\documentclass[3p, a4paper, 12pt]{elsarticle}
\journal{Arxiv}
\pdfoutput=1
\usepackage{color}
\usepackage[usenames,dvipsnames]{xcolor}
\usepackage{bm}
\usepackage{graphicx}
\usepackage{amssymb}
\usepackage{amsmath}
\usepackage{lineno}
\usepackage{multicol}
\usepackage{multirow}
\usepackage{tcolorbox}
\tcbuselibrary{skins, breakable}
\usepackage{subcaption}
\usepackage{booktabs, array, dcolumn}
\usepackage{indentfirst}
\biboptions{numbers,sort&compress}

\newcolumntype{d}{D{.}{.}{2.3}}
\newcolumntype{C}{p}
\usepackage{setspace}
\usepackage{hyperref}
\usepackage{etoolbox}
\usepackage{booktabs}
\hypersetup{
	colorlinks = true,
	citecolor = blue,
	linkcolor=black}

\begin{document}
 
\begin{frontmatter}
 \author[DLUT]{Sitong Tao}
 \author[DLUT]{Fei Han \corref{cor}}

 \cortext[cor]{Corresponding author: hanfei8172@126.com.}
 
  \address[DLUT]{State Key Laboratory of Structural Analysis, Optimization and CAE Software for Industrial Equipment, Department of Engineering Mechanics, International Research Center for Computational Mechanics, Dalian University of Technology, Dalian, 116024, PR China}

\title{A new definition of peridynamic damage for thermo-mechanical fracture modeling}

\begin{abstract}
A thermo-mechanical fracture modeling is proposed to address thermal failure issues, where the temperature field is calculated by a heat conduction model based on classical continuum mechanics (CCM), while the deformation field with discontinuities is calculated by the peridynamic (PD) model. The model is calculated by a CCM/PD alternating solution based on the finite element discretization, which ensures the calculation accuracy and facilitates engineering applications. The original PD model defines damage solely based on the number of broken bonds in the vicinity of the material point, neglecting the distribution of these bonds. To address this limitation, a new definition of the PD damage accounting for both the number of broken bonds and their specific distribution is proposed.  As a result, damage in various directions can be captured, enabling more realistic thermal fracture simulations based on a unified mesh discretization. The effectiveness of the proposed model is validated by comparing numerical examples with analytical solutions. Moreover, simulation results of quasi-static and dynamic crack propagation demonstrate the model's ability to aid in understanding the initiation and propagation mechanisms of complex thermal fractures.
\end{abstract}


\end{frontmatter}

\section{Introduction}
With the rapid development of industry, more and more high-temperature concrete and metal materials are used. However, unpredictable thermal deformation and stress, often resulting from uneven temperature distributions or inconsistent thermal expansion coefficients, can ultimately lead to structural failure. In order to ensure the safety and reliability of the structures, it is necessary to analyze the thermal deformation and thermal stress that may occur in the structures. However, experiment is complex and costly to reproduce the high temperature and high pressure environment, numerical simulation is an alternative approach to help understand the mechanism of thermo-mechanical coupling response of the structure. Therefore, an efficient and accurate numerical simulation method is necessary.
\begin{tcolorbox}[
	enhanced,
	colback=white,  
	colframe=black!75, 
	arc=1mm,         
	boxrule=1pt,   
	fonttitle=\bfseries\large,
	coltitle=black,
	attach boxed title to top center={yshift=-3mm},
	boxed title style={
		colback=white,
		colframe=black!75,
		arc=2mm,
		boxrule=1pt
	},
	after skip=10pt,   
	width=\textwidth,  
	left=5mm,          
	right=5mm,         
	top=5mm,           
	bottom=5mm,        
	]
	\section*{Nomenclature}
	\begin{tabular}{p{2cm}p{12cm}}
		\textbf{Symbol} & \textbf{Definition} \\
		\hline
		PD & Peridynamic \\
		CCM & Classical continuum mechanics \\
		BB-PD & Bond-based peridynamic \\
		$\bm{f}$ & Bond force \\
		$\bm{b}$ & External force \\
		$\bm{\xi}$ & Bond vector\\
		$\bm{u_\xi}$ & The projections of the displacement onto the bond\\
		$\bm{e_\xi}$ & Unit vector of the bond\\
		$\delta$ & Horizon \\
		$H_{\delta(\bm{x})}$ & Neighborhood of point $\bm{x}$ \\
		$\bm{c}$ & Micromodulus tensor \\
		$W$ & Elastic energy density \\
		$\mu$ & Bond-broken variable\\
		$s$ & Bond stretch\\
		$t$ & Computational step\\
		$s_0$ & Critical bond stretch\\
		$G_c$ & Fracture energy \\
		$d$ & PD damage\\
		$\hat{\boldsymbol d}$ & New definition of the PD damage \\
		$d_1, d_2, d_3$ & Damage in the material system \\
		$\boldsymbol e_i^* $ & Basis vector\\
		$d_i^*(\boldsymbol x',\boldsymbol x,t)$ & damage in each direction\\
		$\hat {\boldsymbol T}$ & Temperature variation of the bond\\
		$\boldsymbol T$ & Temperature\\
		$a$ & Micro-expansivity\\
		$\alpha$ & Thermal expansion coefficient\\
		$c$ & Specific heat capacity\\
		$\rho$ & Density\\
		$\bm J$ & Heat flux\\
		$\bm Q$ & Heat source\\
		$\bm k$ & Thermal conductivity\\
		$\bm k_0$ & Reference thermal conductivity\\
		$\bm I$ & Unit matrix\\
		$\theta$ & Integration parameter\\
		$\bm C$ & Heat capacity matrix\\
		$\bm K$ & Heat conduction matrix\\
		$\bm P$ & Temperature load vector\\
	\end{tabular}
\end{tcolorbox}
\noindent
\begin{tcolorbox}[
	enhanced,
	colback=white,  
	colframe=black!75, 
	arc=1mm,         
	boxrule=1pt,   
	fonttitle=\bfseries\large,
	coltitle=black,
	attach boxed title to top center={yshift=-3mm},
	boxed title style={
		colback=white,
		colframe=black!75,
		arc=2mm,
		boxrule=1pt
	},
	after skip=10pt,   
	width=\textwidth,  
	left=5mm,          
	right=5mm,         
	top=5mm,           
	bottom=5mm,        
	]
	\begin{tabular}{p{2cm}p{12cm}}
		$n$ & The number of total finite elements\\
		$h$ & Heat convection coefficient\\
		$\boldsymbol \phi_{a}$ & Temperature on boundary\\
		$\bm N$ & Matrix of shape function\\
		$\bm H$ & Matrix of differential operators\\
		$\hat{\bm K}$ & Total stiffness matrix\\
		$\bm E$ & Young's modulus\\
		$\mu$ & Poisson's ratio\\
		$G$ & Fracture energy\\
	\end{tabular}
\end{tcolorbox}\noindent
CCM is widely used for continuous thermo-mechanical problems, but its partial differential governing equation cannot handle discontinuous issues such as thermal fracture. To address this, several numerical methods have been developed, including extended finite element method (XFEM) \cite{XFEM}, phase-field fracture method (PFM) \cite{phasefield}, and discrete element method (DEM) \cite{DEM}. XFEM is effective for simulating discontinuous problems such as interfaces and crack propagation. Jaskowiec et al. used XFEM for three-dimensional numerical thermo-mechanical modeling of a laminated structure with a very thin inner layer \cite{jaskowiec}. Kumar et al. employed XFEM for a thermo-mechanical fracture analysis of porous functionally graded cracked plates \cite{kumar}. PFM is used for simulating structural damage. Badnava et al. used PFM to simulate brittle fracture and thermal cracks in two-dimensional (2D) and three-dimensional (3D) continua \cite{badnava2018}. Zhou et al. presented a novel coupled thermo-mechanical PFM for concrete at high temperatures \cite{zhou2024}. DEM can reproduce macroscopic behavior comparable to laboratory tests and monitor microscopic variations in the failure process. Sun et al. presented a low-temperature thermo-mechanical coupling modeling framework to simulate frost crack evolution in rock masses using the finite-discrete element method (FDEM) \cite{sun2022}. Although the above methods can solve the thermo-mechanical coupling problem, it is still a challenge to deal with the initiation and propagation of complex multi-cracks.

Silling introduced a novel non-local continuum model known as the peridynamic (PD) model \cite{silling2000}. PD model has advantages for complex multi-cracks with unknown a priori locations. Unlike the partial differential governing equations used in CCM, PD model employs integro-differential governing equations, making it suited for simulation structural fracture. PD naturally simulates crack initiation and propagation without requiring any predefined crack growth criteria. Subsequently, a PMB constitutive model of PD was introduced by Silling \cite{silling_meshfree_2005}. This particular PD model is termed the bond-based peridynamic (BB-PD) model, wherein the Poisson’s ratio remains constrained to a fixed value. Recognizing this limitation, Silling introduced a mathematical construct known as `state' in 2007, thereby presenting two variations: the ordinary state-based peridynamic (OSB-PD) model and the non-ordinary state-based peridynamic (NOSB-PD) model \cite{silling2007}. Notably, the BB-PD model can be viewed as a specialized instance within the broader framework of the state-based peridynamic model \cite{silling_linearized_2010}. Additionally, PD model demonstrates applicability in addressing thermal conduction and thermal fracture problems. Bobaru and Duangpanya introduced a PD formulation for transient heat conduction in solids with discontinuities \cite{bobaru2012}. Oterkus et al. derived OSB-PD heat conduction equations \cite{oterkus2014a}. Based on the above works, PD can be used to solve thermo-mechanical problems. Oterkus et al. presented a fully coupled PD thermo-mechanical framework \cite{oterkus2014}. PD's inherent ability to simulate crack initiation and propagation has made it practical for engineering applications. Wang et al. developed a thermo-mechanical BB-PD model to simulate thermal cracking processes in concrete exposed to fire scenarios \cite{wang2024}. Zang et al. introduced a fully coupled thermo-mechanical PD model for rock fracturing under blast loading, accounting for initial pore damage \cite{zhang2024}. Cheng et al. presented a thermo-mechanical PD model to investigate damage in engineered cementitious composite-concrete bonding specimens at high temperatures \cite{cheng2024}. While these studies effectively simulated thermal fracture, selecting the appropriate micro-conductivity remains a challenge. Sun et al. presented a novel computational framework for analyzing thermal fracture in brittle solids by coupling PD and CCM \cite{sun2021}. However, the original PD model defines damage solely based on the number of broken bonds, but neglects their spatial distribution, thereby introducing inaccuracies in damage quantification. To address this limitation, we introduce a novel PD damage formulation that considers both the quantity and spatial distribution of broken bonds. As a result, damage in various directions can be captured, enabling the simulation of thermal fracture based on a unified mesh discretization framework.

The structure of the remainder of this paper is organized as follows: Section \ref{sec:sec2} reviews the fundamental formulations of the BB-PD model and presents the new definition of the PD damage within this framework. Section \ref{sec:sec3} revisits the thermo-mechanical PD model and establishes a framework for integrating the new definition of the PD damage into the model. Section \ref{sec:sec4} details the finite element spatial discretization, as well as the time discretization, of the proposed framework. Section \ref{sec:sec5} demonstrates the effectiveness of the proposed model through four examples. Section \ref{sec:sec6} concludes with remarks summarizing the findings and contributions of this paper.

\section{A new definition of the PD damage} \label{sec:sec2}
In this chapter, we first review the BB-PD model and the original definition of the PD damage. Then, we elucidate the necessity and specific formula of the new definition of the PD damage. It should be noted that the new definition of the PD damage presented in this paper is applicable to the NOSB-PD model and the OSB-PD model, with the BB-PD model being introduced in detail as a special case.
\subsection{A review of the BB-PD model and the PD damage}
PD model assumes that each material point $\boldsymbol x$ has its own neighborhood $\mathcal{H}_\delta(\boldsymbol x)$ and interacts with points $\boldsymbol x^{\prime}$ located in $\mathcal{H}_\delta(\boldsymbol x)$. The equilibrium equation can be expressed as follows \cite{silling2010}:
\begin{align}
	\int_{\mathcal{H}_{\delta}(\boldsymbol x)}\boldsymbol f(\boldsymbol x^{\prime},\boldsymbol x)\mathrm{d} V_{\boldsymbol x^{\prime}}+\boldsymbol b(\boldsymbol x)=0\quad\forall\:\boldsymbol x,\boldsymbol x^{\prime}\in\Omega   
\end{align}
where $H_\delta(\boldsymbol x)$ denotes the neighborhood of $\boldsymbol{x}$ with a horizon of $\delta$; $\boldsymbol b(\boldsymbol x)$ signifies the external force acting on point $\boldsymbol{x}$, and $\boldsymbol f(\boldsymbol x^{\prime},\boldsymbol x)$ represents a pairwise force function. A potential constitutive model relating force to relative displacement for linear elasticity and small deformations can be expressed as follows:
\begin{align}
	\boldsymbol f(\boldsymbol x^{\prime},\boldsymbol x)=\:\frac{1}{2}\left(\boldsymbol c(\boldsymbol x,|\boldsymbol \xi|)+\boldsymbol c(\boldsymbol x^{\prime},|\boldsymbol \xi|)\right)\left(\boldsymbol u_{\boldsymbol \xi}(\boldsymbol x^{\prime})-\boldsymbol u_{\boldsymbol \xi}(\boldsymbol x)\right)\boldsymbol e_{\boldsymbol \xi} 
\end{align}
where $\boldsymbol u_{\boldsymbol \xi}(\boldsymbol x^{\prime})$ and $\boldsymbol u_{\boldsymbol \xi}(\boldsymbol x)$ represent the projections of the displacement at point $\boldsymbol x$ and $\boldsymbol x^{\prime}$ onto the bond, respectively; $\boldsymbol e_{\boldsymbol \xi}$ is the unit vector of bond $\boldsymbol \xi$;  $\boldsymbol c(\boldsymbol x,|\boldsymbol \xi|)$ and $(\boldsymbol c(\boldsymbol x^{\prime},|\boldsymbol \xi|))$ denote micro-modulus functions of bond $\boldsymbol \xi$. In this paper, we focus on homogeneous materials (i.e.,$\boldsymbol c(\boldsymbol x,|\boldsymbol \xi|)=\boldsymbol c(\boldsymbol x^{\prime},|\boldsymbol \xi|)=\boldsymbol c^{0}(|\boldsymbol \xi|)$). The elastic energy density can be expressed as follows \cite{LUBINEAU20121088}:
\begin{align}
	W(\boldsymbol x)&=\frac{1}{4}\int_{\mathcal{H}_{\delta}(\boldsymbol x)}\boldsymbol c^{0}(|\boldsymbol \xi|)\left(\boldsymbol u_{\boldsymbol \xi}(\boldsymbol x^{\prime})-\boldsymbol u_{\boldsymbol \xi}(\boldsymbol x)\right)^{2}\mathrm{d} V_{\boldsymbol x^{\prime}}
\end{align}

To capture crack initiation and propagation, the PD model requires a bond failure criterion to trigger bond failure. Bond failure can be tracked using a history-dependent function $\mu$, defined as follows:
\begin{align}
	\mu(\boldsymbol x',\boldsymbol x,t)=\begin{cases}\:1& s\:<s_0\\\:0&\text{otherwise}\end{cases} 
\end{align}
where $s$ is the bond stretch; $t$ is the computational step; and $s_0$ denotes the critical bond stretch. The relationship between the bond stretch $s$ and displacement can be formulated as follows:
\begin{align}
	s=\frac{|\boldsymbol u(\boldsymbol x')-\boldsymbol u(\boldsymbol x)+\boldsymbol \xi|-|\boldsymbol \xi|}{|\boldsymbol \xi|}
\end{align}

It's worth noting that, PD model employs a scalar $d(\boldsymbol x,t)$ to describe damage at a point $\boldsymbol x$: 
\begin{align}\label{equ:damage}
	d(\boldsymbol x,t)=1-\frac{\int_{\mathcal{H}_\delta(\boldsymbol x)}\mu(\boldsymbol x',\boldsymbol x,t)\mathrm{d} V_{\boldsymbol x'}}{\int_{\mathcal{H}_\delta(\boldsymbol x)}\mathrm{d} V_{\boldsymbol x'}}  
\end{align}

\subsection{A new definition of the PD damage based on the spatial distribution of the broken bonds}
\begin{figure}[!h]
\centering
\includegraphics[width=4in]{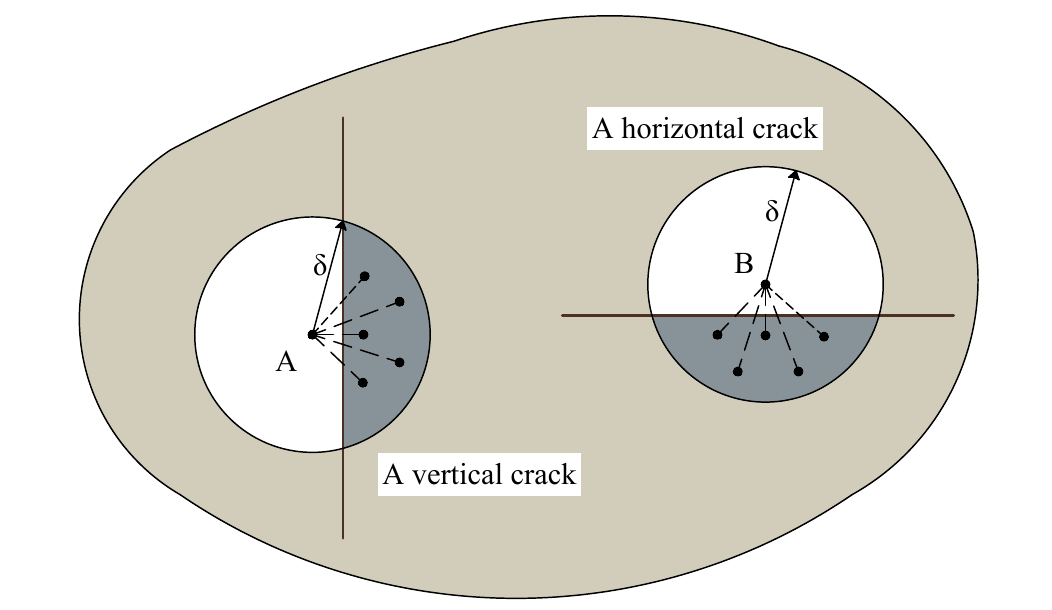}
\caption{Comparison of the spatial distribution of the broken bonds between points $A$ and $B$.}
\label{fig:fig1}
\end{figure}

As demonstrated in Eq. (\ref{equ:damage}), the original PD damage is calculated by the proportion of broken bonds to total bonds. While the original PD damage can indeed characterize the degradation of materials, it is difficult to accurately portray the specific bond failure distribution within the neighborhood. To illustrate, consider the scenario depicted in Fig. \ref{fig:fig1}, where two differently oriented cracks traverse the neighborhood in a two-dimensional plane. The spatial distribution of the broken bonds differ between the vertical crack and the horizontal crack. However, when Eq. (\ref{equ:damage}) is employed to compute the damage values for material points $\boldsymbol A$ and $\boldsymbol B$, these two distinct scenarios may have identical damage values ($d(\boldsymbol A) = d(\boldsymbol B) = 0.5$). It underscores the limitation of the original PD damage in capturing the specific spatial distribution of the broken bonds. On the other hand, the original PD damage establish a homogenized mapping mechanism between microscopic bond failure and macroscopic damage fields through statistical averaging approaches. However, when addressing multi-physics coupling simulations (e.g., thermo-mechanical or hygro-mechanical interactions), the macroscopic damage field necessitates explicit consideration of anisotropic damage characteristics. This requirement exposes an intrinsic limitation of original PD damage — its inherent isotropy assumption fundamentally restricts the characterization of direction-dependent damage evolution patterns. 

To address the aforementioned limitation in the PD damage, it is necessary to implement a new definition of the PD damage based on bonds distribution. In the new definition, the PD damage can be represented as $\hat{\boldsymbol d}$ rather than a scalar to accurately describe damage in various directions. 

\begin{align}
	 \hat{\boldsymbol d} = 
\begin{pmatrix}
	d_{1} & 0 & 0 \\
	0 & d_{2} & 0 \\
	0 & 0 & d_{3}
\end{pmatrix}
\end{align}
where $d_1$, $d_2$ and $d_3$ are components defined in the material coordinate system. For isotropic materials, $d_{1}$, $d_{2}$, and $d_{3}$ represent the damage in the directions of the x, y, and z axes, respectively. And for anisotropic  materials, $d_{1}$, $d_{2}$, and $d_{3}$ represent damage in three main directions. However, due to the symmetry of the PD domain, despite dividing the bond among different directions, for the same PD neighborhood, $d_1$, $d_2$ and $d_3$ are almost equal. Therefore, it remains challenging to distinguish damage among different directions.

\begin{figure}[h!]
	\centering
	\subcaptionbox{Four unit basis vectors in the 2D case.}{
		\includegraphics[width=0.45\linewidth]{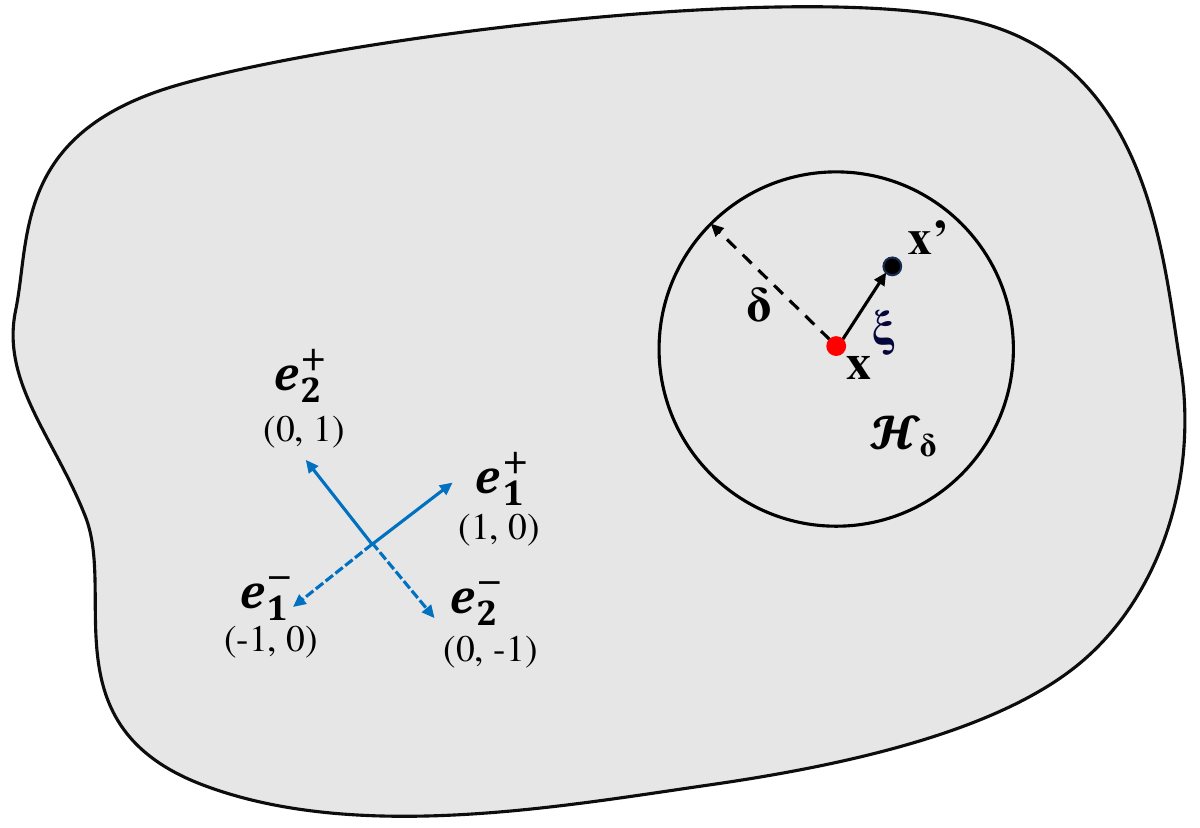}
	}
	\subcaptionbox{Six unit basis vectors in the 3D case.}{
		\includegraphics[width=0.45\linewidth]{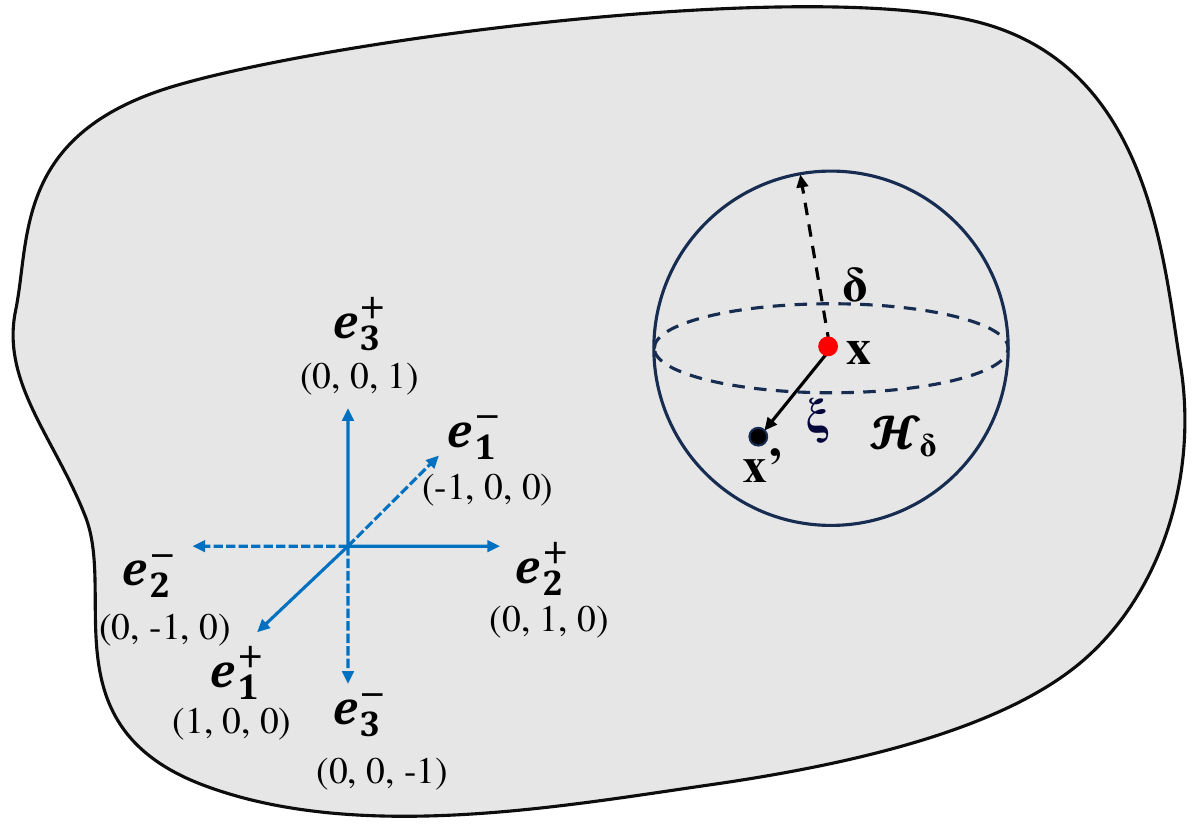}
	}
	\caption{Basis vectors $\boldsymbol e_i^*$ in 2D and 3D cases.}
	\label{fig:fig2}
\end{figure}

\begin{figure}[h!]
	\centering
		\subcaptionbox{Crack path with an angle of $\alpha$ to the principal axes of material.}{
		\includegraphics[height=0.35\linewidth]{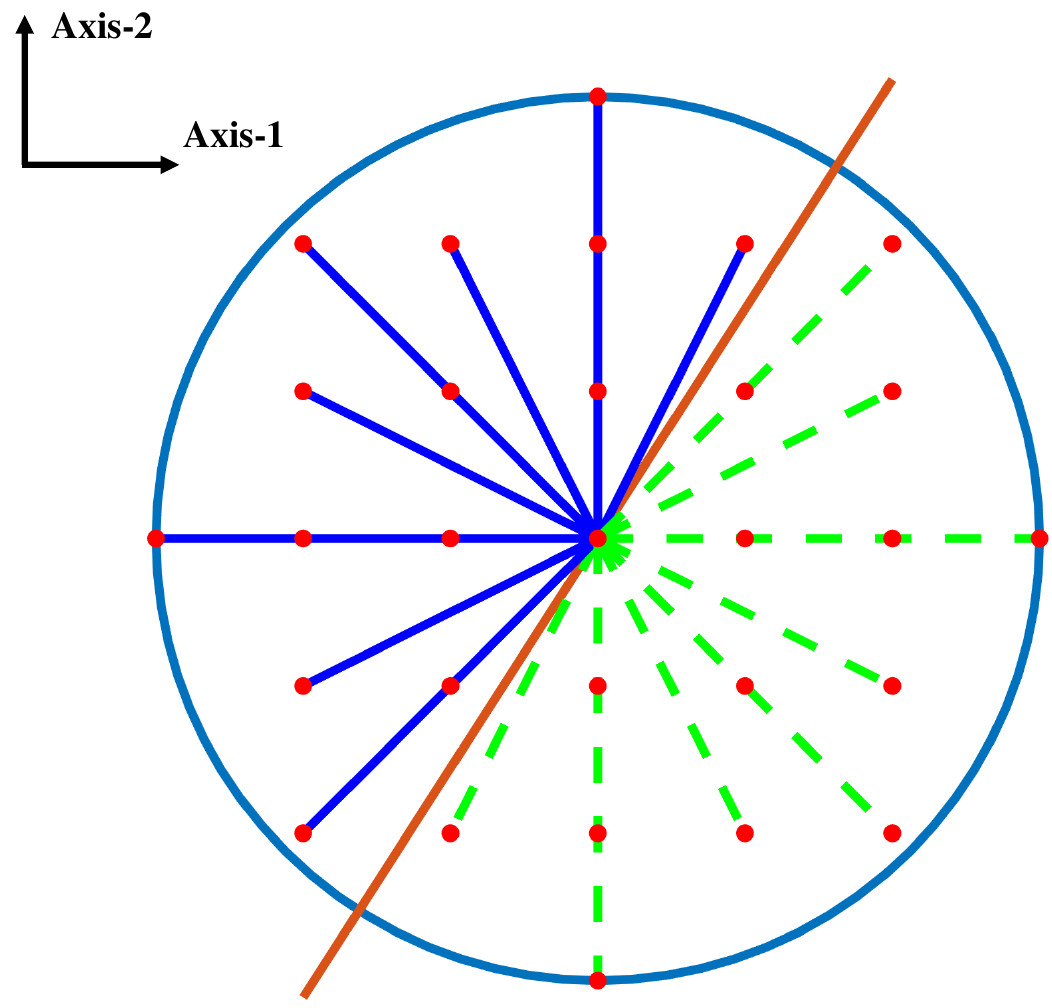}
	}
	\subcaptionbox{The classical damage and the redefined damage.}{
		\includegraphics[height=0.35\linewidth]{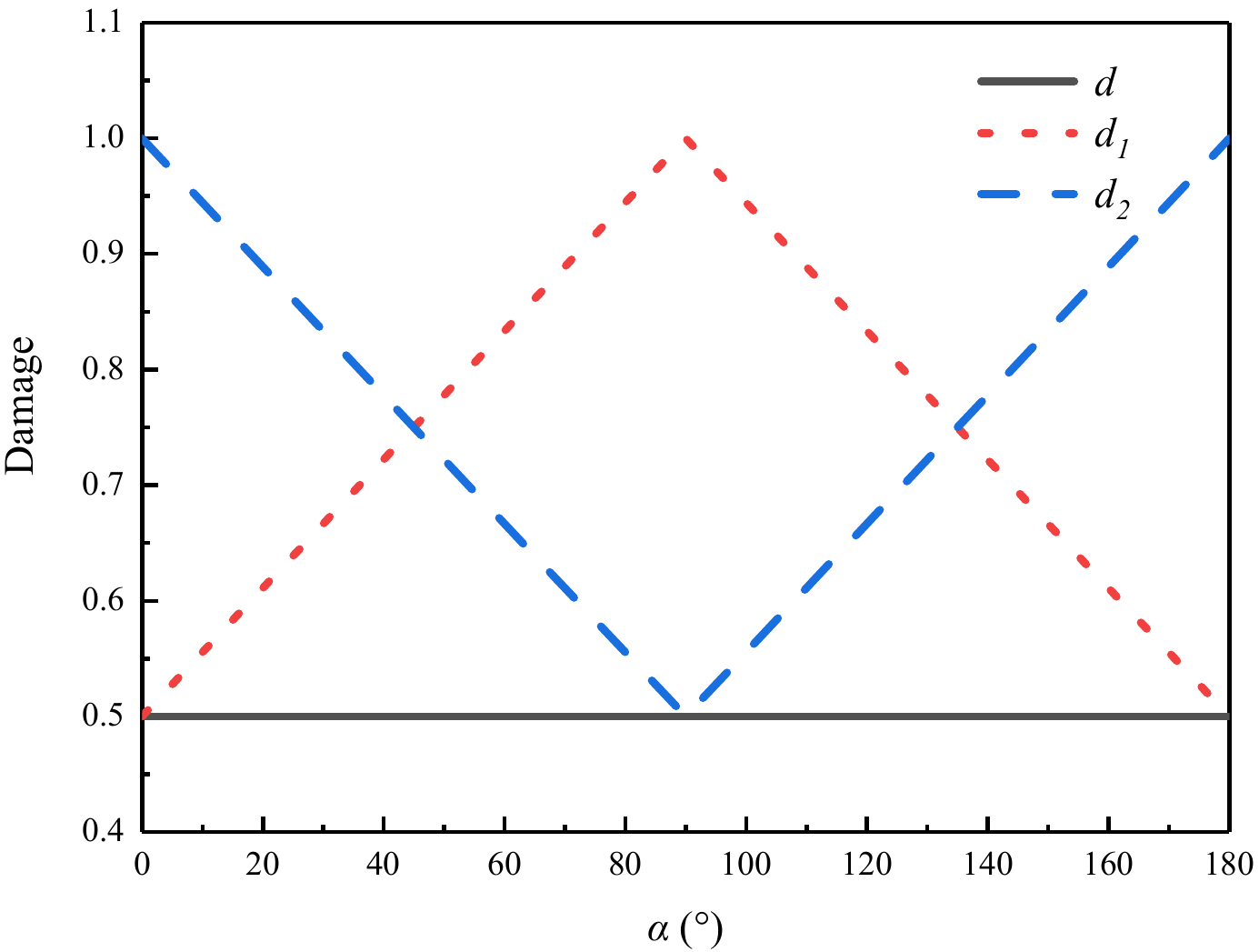}
	}
	\caption{Comparison of damage at different crack angles.}
	\label{fig:figbond}
\end{figure}
To distinguish $d_1$, $d_2$ and $d_3$, some improvements have been made to present a new definition of the PD damage. As shown in Fig. \ref{fig:fig2}, unit basis vectors $\boldsymbol e_i^* (*=+,-)$ along material principal directions are defined in both 2D and 3D cases. For 2D cases, $i=1,2$ and for 3D cases, $i=1,2,3$. To distinguish the positions of each bond $\boldsymbol\xi$ in the neighborhood $\mathcal{H}_\delta$, a scalar function $\nu_i^*(\boldsymbol x',\boldsymbol x)$ is defined as follow:
\begin{align}
	\nu_i^*(\boldsymbol x',\boldsymbol x)=
	\begin{cases}\:1& \boldsymbol e_i^*\cdot\boldsymbol \xi\:>0\\\:0&\text{otherwise}
	\end{cases} 
	\label{equ:nu}
\end{align}

Based on Eqs. (\ref{equ:damage}) and (\ref{equ:nu}) the following formula can be obtained:
\begin{align}
	d_i^*(\boldsymbol x',\boldsymbol x,t)=1-\frac{\int_{\mathcal{H}_\delta(\boldsymbol x)}\nu_i^*(\boldsymbol x',\boldsymbol x)\mu(\boldsymbol x',\boldsymbol x,t)\mathrm{d} V_{\boldsymbol x'}}{\int_{\mathcal{H}_\delta(\boldsymbol x)}\nu_i^*(\boldsymbol x',\boldsymbol x)\mathrm{d} V_{\boldsymbol x'}}  
\end{align}
where $d_i^*(\boldsymbol x',\boldsymbol x,t)$ ($0\leq d_i^*(\boldsymbol x',\boldsymbol x,t)\leq 1$) denotes damage in each direction $i$, $*$. For 2D cases, $i=1, 2$ and for 3D cases, $i=1, 2, 3$.

The above formula means, if asymmetric bond failure occur within the domain, different quadrants may exhibit varying damage values. Through comparison, the quadrant with the maximum damage is selected as part of damage $\hat{\boldsymbol d}$. For 2D cases, the new definition of the PD damage can be written as:
\begin{align}\label{equ:d2}
	\hat{\boldsymbol d}=
	\begin{pmatrix}
	d_1 & 0\\
	0 & d_2
	\end{pmatrix} = 	
	\begin{pmatrix}
	max\{d_1^+,d_1^-\} & 0\\
	0 & max\{d_2^+,d_2^-\}
	\end{pmatrix} 
\end{align}
This means that the maximum vector should be selected among all possible damage vectors. Similarly, for 3D cases, the new definition of the PD damage can be written as:
\begin{align}\label{equ:d3}
	\hat{\boldsymbol d}=
	\begin{pmatrix}
		d_1 & 0 & 0\\
		0 & d_2 & 0\\
		0 & 0 & d_3\\
	\end{pmatrix} = 	
	\begin{pmatrix}
		max\{d_1^+,d_1^-\} & 0 & 0\\
		0 & max\{d_2^+,d_2^-\} & 0 \\
		0 & 0 & max\{d_3^+,d_3^-\}\\
	\end{pmatrix} 
\end{align}

As shown in Fig. \ref{fig:fig1}, for point $A$, the new definition of the PD damage $\hat{\boldsymbol d}=
\begin{pmatrix}
	1 & 0 \\
	0 &0.5\\
\end{pmatrix}$, whereas for point $B$, $\hat{\boldsymbol d}=
\begin{pmatrix}
0.5 & 0 \\
0 & 1\\
\end{pmatrix}$. This means that the new definition of the PD damage varies depending on different bond failure distributions within the PD domain. In order to further clarify the effectiveness of the new definition of the PD damage in characterizing cracks, as depicted in Fig. \ref{fig:figbond}(a), consider a crack surface that intersects the horizontal plane. Assume that all bonds crossing the crack surface are broken. The angle between this crack surface and the principal axes of material is denoted as $\alpha$. Broken bonds are indicated by green dashed lines, while intact bonds are represented by blue solid lines. Furthermore, as illustrated in Fig. \ref{fig:figbond}(b), red line represents $d_1$, blue line represents $d_2$ and black line represents the original PD damage. As the angle $\alpha$ changes, the new definition of the PD damage also varies accordingly. When the angle is $0^\circ$, $d_2$ is maximum and $d_1$ is minimum. When the angle is $90^\circ$, $d_1$ is maximum and $d_2$ is minimum.

The subsequent chapters introduce the applications and benefits of the new definition of the PD damage in modeling thermal fracture.

\section{A new definition of the PD damage for modeling thermal fracture}\label{sec:sec3}

\subsection{An improved thermo-mechanical PD model}\label{sec:sec3.1}
The PD equilibrium equation with temperature can be expressed as follows:
\begin{align}
	\int_{\mathcal{ H}_{\delta}(\boldsymbol x)}\boldsymbol f(\boldsymbol x^{\prime},\boldsymbol x,\hat {\boldsymbol T})\mathrm{d} V_{\boldsymbol x^{\prime}}+\boldsymbol b(\boldsymbol x)=0\quad\forall\:\boldsymbol x,\boldsymbol x^{\prime}\in\Omega   
\end{align}
The bond force can be divided into two parts:
\begin{align}\label{equ:bond1}
	\boldsymbol f(\boldsymbol x',\boldsymbol x,\hat{\boldsymbol T})=\hat{\boldsymbol f}(\boldsymbol x',\boldsymbol x,\hat{\boldsymbol T})-\hat{\boldsymbol f}(\boldsymbol x,\boldsymbol x',\hat{\boldsymbol T})
\end{align}
where $\hat{\boldsymbol f}(\boldsymbol x',\boldsymbol x,\hat{\boldsymbol T})$ and $\hat{\boldsymbol f}(\boldsymbol x,\boldsymbol x',\hat{\boldsymbol T})$ are bond forces of point $\boldsymbol x'$ over point $\boldsymbol x$ and $\boldsymbol x$ over point $\boldsymbol x'$; $\hat{\boldsymbol T}$ is the temperature variation of the bond. A possible constitutive equation can be written as \cite{kilic2010}:
\begin{align}\label{equ:bond2}
	\hat{\boldsymbol f}(\boldsymbol x',\boldsymbol x,\hat{\boldsymbol T})=\frac{1}{2}\boldsymbol c(\boldsymbol x,|\boldsymbol \xi|)(\boldsymbol u_{\boldsymbol \xi}(\boldsymbol x')-\boldsymbol u_{\boldsymbol \xi}(\boldsymbol x)-a(\boldsymbol x)\hat{\boldsymbol T}(\boldsymbol x',\boldsymbol \xi))\boldsymbol e_{\boldsymbol \xi}
\end{align}
where $a(\boldsymbol x)$ denotes micro-expansivity at point $\boldsymbol x$,for homogeneous materials, $a(\boldsymbol x)=a(\boldsymbol x')=a^0$.
Substitute Eq. (\ref{equ:bond2}) into Eq. (\ref{equ:bond1}):
\begin{equation}
\begin{aligned}
	\boldsymbol f(\boldsymbol x^{\prime},\boldsymbol x,\hat{\boldsymbol T})&=\:\frac{1}{2}\left(\boldsymbol c(\boldsymbol x,|\boldsymbol \xi|)+\boldsymbol c(\boldsymbol x^{\prime},|\boldsymbol \xi|)\right)\left(\boldsymbol u_{\boldsymbol \xi}(\boldsymbol x^{\prime})-\boldsymbol u_{\boldsymbol \xi}(\boldsymbol x)\right)\boldsymbol e_{\boldsymbol \xi}
	\\&-\frac{1}{2}\left(\boldsymbol b(\boldsymbol x,|\boldsymbol \xi|)+\boldsymbol b(\boldsymbol x^{\prime},|\boldsymbol \xi|)\right)\hat{\boldsymbol T}(\boldsymbol x,\boldsymbol \xi)\boldsymbol e_{\boldsymbol \xi}
\end{aligned}
\end{equation}
where $\boldsymbol b(\boldsymbol x,|\boldsymbol \xi|)$ is the thermal modulus, written as:
\begin{align}
	\boldsymbol b(\boldsymbol x,\boldsymbol \xi)= a(\boldsymbol x)\boldsymbol c(\boldsymbol x,|\boldsymbol \xi|) = a^0c^0(|\boldsymbol \xi|)
\end{align}

In this paper, we focus on homogeneous materials. (i.e.,$\boldsymbol b(\boldsymbol x,|\boldsymbol \xi|)=\boldsymbol b(\boldsymbol x^{\prime},|\boldsymbol \xi|)=\boldsymbol b^{0}(|\boldsymbol \xi|)$)

Consider a uniform and isotropic solid containing a heat source and undergoing heat exchange with its surrounding medium. Study the distribution and variations of temperature within the solid. This analysis is grounded in the principles of the energy conservation equation:
\begin{align}
\frac{\mathrm{d}}{\mathrm{d}t}\iiint_{R}c\rho \boldsymbol{T}\mathrm{d}V =-\iiint_{R}\nabla\cdot\boldsymbol{J}\mathrm{d}V +\iiint_{R}\boldsymbol Q\mathrm{d}V.
\end{align}
where $c$ is the specific heat capacity; $\rho$ is the density; $\boldsymbol{T}$ is the temperature; $\boldsymbol{J}$ is the heat flux; $t$ is the time, $\boldsymbol Q$ is the heat source. The above equation can be simplified as:
\begin{align}
	\rho c\boldsymbol{\dot{T}}+\nabla\cdot\boldsymbol{J}=\boldsymbol Q
\end{align}

According to the Fourier law,
\begin{align}
	\boldsymbol{J}=-\boldsymbol k\nabla{\boldsymbol{T}}
\end{align}
where $\boldsymbol k$ is the thermal conductivity.

\subsection{Anisotropic thermal conductivity for modeling thermal fracture}
According to Section \ref{sec:sec3.1}, in the thermo-mechanical model, the temperature field and deformation field can interact with each other. Changes in temperature, whether increasing or decreasing, affect the material's thermal deformation. The bond failure can be calculated through deformation fields. Progressive accumulation of microscopic bond failures induces macroscopic damage which can locally characterize the degradation of thermal conductivity.

\begin{figure}[h!]
	\centering
	\includegraphics[width=4.5in]{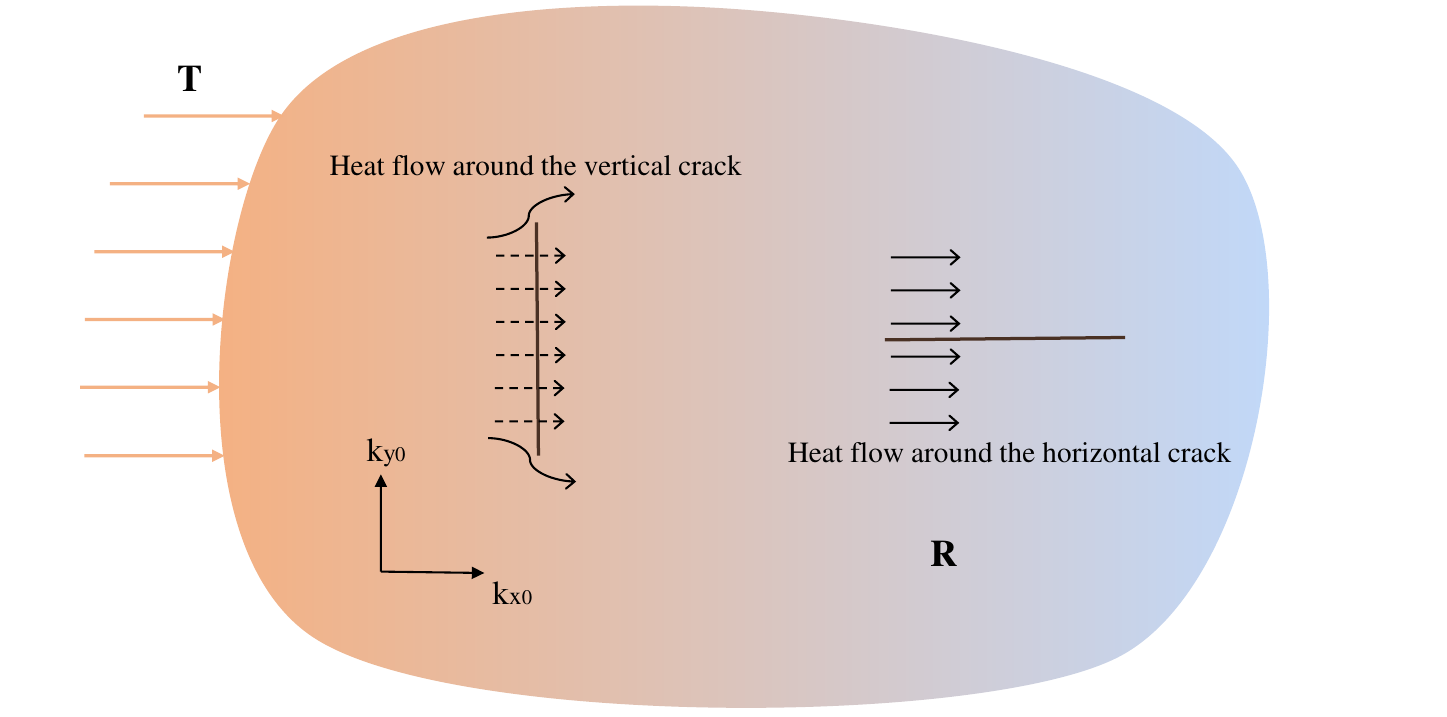}
	\caption{Heat flow around the vertical crack and horizontal crack.}
	\label{fig:fig3}
\end{figure}

As shown in Fig. \ref{fig:fig3}, a fixed temperature $\boldsymbol T$ is applied to the left boundary of the solid. When considering heat flow through both horizontal and vertical cracks, it is observed that the vertical crack significantly hinders heat flow, while the horizontal crack has a weaker impact. It demonstrates that cracks oriented in different directions can hinder heat flow to varying degrees. Therefore, the reduction in thermal conductivity cannot solely be attributed to the original PD damage, as described by Eq. (\ref{equ:damage}), and must also consider the direction and extent of cracking.

A suitable way to define thermal conductivity $\boldsymbol k$ by PD damage is as follows:
\begin{align}\label{equ:k}
	\boldsymbol k = \left(\boldsymbol I - \hat{\boldsymbol d} \right) \boldsymbol k_0
\end{align}
where $\boldsymbol k_0$ is the reference thermal conductivity; $\boldsymbol I$ is the unit matrix; $\hat{\boldsymbol d}$ is defined in Eqs. (\ref{equ:d2}) and (\ref{equ:d3}).
This formula reflects the anisotropic effect of damage on thermal conductivity, with different degrees of degradation in different directions.
\section{Numerical algorithm}\label{sec:sec4}
\subsection{A unified finite element discretization for thermo-mechanical crack propagation}
\begin{figure}[h!]
	\centering
	\includegraphics[width=0.98\linewidth]{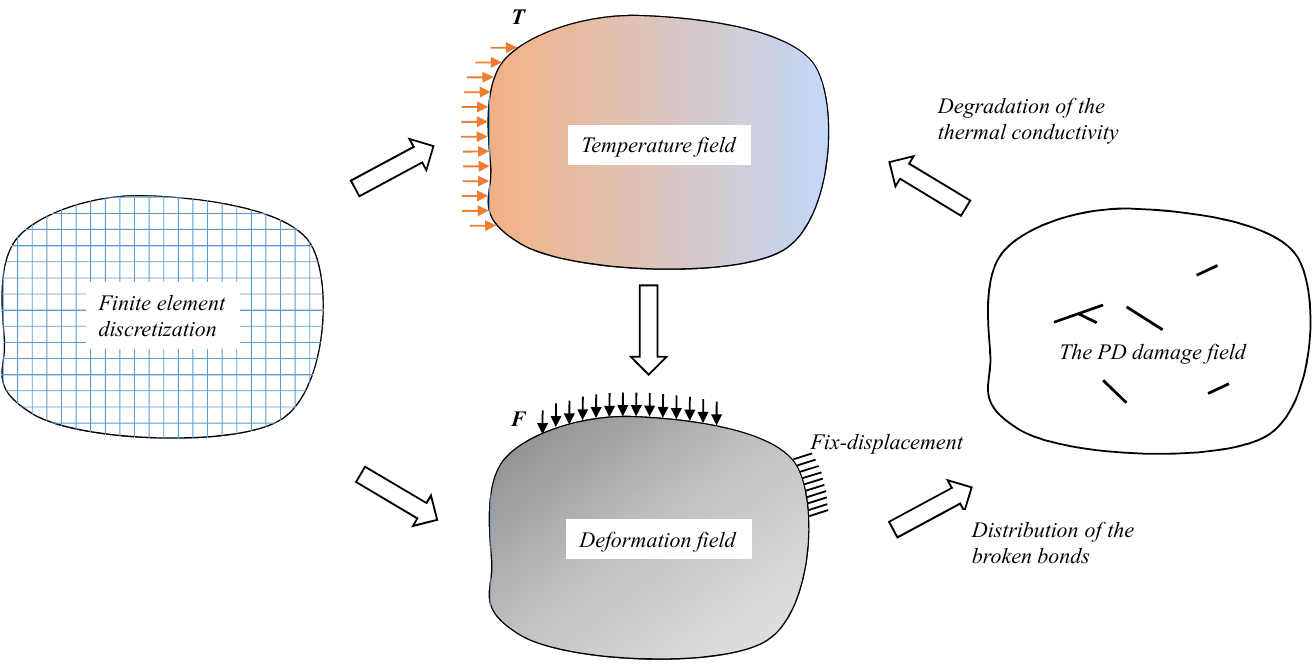}
	\caption{A unified finite element discretization for thermo-mechanical crack propagation.}
	\label{fig:fig4}
\end{figure}

In previous study \cite{sun2021}, temperature fields were computed by a local model based on the finite element discretization, while deformation fields with discontinuities were computed by PD model based on an element-free discretization. In this paper, we unifies the discretization of both model through a shared mesh system. Fig. \ref{fig:fig4} illustrates the proposed computational framework, where finite element discretization of the computational domain is implemented during the initialization phase. The computational procedure initiates with the computation of the temperature field. Sequentially, temperature field is incorporated into the PD model to obtain the deformation field. The deformation field is used to determine bond failures. Progressive accumulation of bond failures induces micro-cracks nucleation. Micro-cracks coalesce through damage bands to form macro-cracks. The PD damage defined in this paper enables characterization of degradation in thermal conductivity, affecting the distribution of the temperature field.

\subsection{Time and spatial discretization}
The time discretization of the temperature field is as follows:
\begin{align}
	\boldsymbol T(n+\theta\Delta t)&=(1-\theta)\boldsymbol T_n+\theta \boldsymbol T_{n+1}\\\dot{\boldsymbol T}(n+\theta\Delta t)&=(\boldsymbol T_{n+1}-\boldsymbol T_{n})/\Delta t 
\end{align}
where $\Delta t$ is the time increment; $\theta$ is the integration parameter; different values of $\theta$ correspond to different differential formulas. $\boldsymbol T_{n+1}$ and $\boldsymbol T_n$ are temperatures at time step $n+1$ and $n$.

The governing equation for temperature field computation can be expressed in the following canonical finite element form:
\begin{align}\label{equ:temp}
	\boldsymbol C\dot{\boldsymbol T}+\boldsymbol K\boldsymbol T=\boldsymbol P
\end{align}
where $\boldsymbol C$ is the heat capacity matrix; $\boldsymbol K$ is the heat conduction matrix; $\boldsymbol T$ is the temperature vector; $\boldsymbol P$ is the temperature load vector; $\dot{\boldsymbol T}$ is the derivative vector of node temperature with respect to time.

The heat conduction matrix $\boldsymbol K$, the heat capacity matrix $\boldsymbol C$ and the temperature load vector $\boldsymbol P$ can be written as:
\begin{equation}
\begin{aligned}
	\boldsymbol K&=\sum_{i=1}^n\int_{V_i}(\boldsymbol H \boldsymbol N_i(\boldsymbol x)\boldsymbol R_i)^T\boldsymbol k(\boldsymbol H \boldsymbol N_i(\boldsymbol x)\boldsymbol R_i)\mathrm{d}V_{\boldsymbol x }\\ &+\sum_{i=1}^n\int_{S_{i}^3}h(\boldsymbol N_i(\boldsymbol x)\boldsymbol R_i)^T(\boldsymbol N_i(\boldsymbol x)\boldsymbol R_i)\mathrm{d}S_x^3\\
	\boldsymbol C&=\sum_{i=1}^n\int_{V_i}\rho c(\boldsymbol N_i(\boldsymbol x)\boldsymbol R_i)^T(\boldsymbol N_i(\boldsymbol x)\boldsymbol R_i)\mathrm{d}V_{\boldsymbol x } \\
	\boldsymbol P&=\sum_{i=1}^n\int_{V_i}\rho \boldsymbol Q(\boldsymbol x)(\boldsymbol N_{i}(\boldsymbol x)\boldsymbol R_i)^T\mathrm{d}V_{\boldsymbol x }\\&-\sum_{i=1}^n\int_{S_{i}^2}\boldsymbol q(\boldsymbol x)(\boldsymbol N_{i}(\boldsymbol x)\boldsymbol R_i)^T\mathrm{d}S_x^2\\&+\sum_{i=1}^n\int_{S_{i}^3}h\boldsymbol \phi_{a} (\boldsymbol N_{i}(\boldsymbol x)\boldsymbol R_i)^T\mathrm{d}S_x^3
\end{aligned}
\end{equation}
where $n$ is the number of total finite elements; $h$ is the heat convection coefficient corresponding to the boundary $S^{3}$; $\boldsymbol \phi_{a}$ is the temperature on boundary $S^{3}$; $\boldsymbol q$ is the heat flux density corresponding to the boundary $S^{2}$; $\boldsymbol N$ denotes the matrix of shape function; $\boldsymbol H$ denotes the matrix of differential operators; $\boldsymbol Q$ is the heat source.

The finite element spatial discretization of PD model can be written as the following formula \cite{liu2022}:
\begin{align}\label{equ:force}
	\hat {\boldsymbol K}\boldsymbol d=\boldsymbol F
\end{align} 
where $\hat {\boldsymbol K}$ is the total stiffness matrix; $\boldsymbol d$ is the displacement vector; $\boldsymbol F$ is the external load force vector. 

The total stiffness matrix $\hat {\boldsymbol K}$ and the external load force vector $\boldsymbol{F}$ can be written as: 
\begin{equation}
\begin{aligned}
	\hat {\boldsymbol K}&= \frac12\sum_{i=1}^n\sum_{j=1}^{\hat h(x)}\int_{V_i}\int_{V_x^j}c^0(|\boldsymbol \xi|)(\boldsymbol N_j(\boldsymbol x')\boldsymbol R_j-\boldsymbol N_i(\boldsymbol x)\boldsymbol R_i)^\mathrm{T}\frac{\boldsymbol \xi\otimes\boldsymbol \xi}{|\boldsymbol \xi|^2}(\boldsymbol N_j(\boldsymbol x^{\prime})\boldsymbol R_j-\boldsymbol N_i(\boldsymbol x)\boldsymbol R_i)\mathrm{d}V_{\boldsymbol x^{\prime}}\mathrm{d}V_{\boldsymbol x }
	\\
	\boldsymbol{F}&=\sum_{i=1}^n\int_{V_i}(\boldsymbol N_i(\boldsymbol x)\boldsymbol R_i)^\mathrm{T}\boldsymbol b(\boldsymbol x)\mathrm{d}V_x +\sum_{i=1}^n\int_{S_i}(\boldsymbol N_i(\boldsymbol x)\boldsymbol R_i)^\mathrm{T}\overline{\boldsymbol F}(\boldsymbol x)\mathrm{d}S_x\\
	&+ \frac12\sum_{i=1}^n\sum_{j=1}^{\hat h(x)}\int_{V_i}\int_{V_x^j}b^0(|\boldsymbol \xi|)(\boldsymbol N_{j}(\boldsymbol x^{\prime})\boldsymbol R_{j}-\boldsymbol N_{i}(\boldsymbol x)\boldsymbol R_{i})^{\mathrm{T}}\frac{\boldsymbol \xi}{|\boldsymbol \xi|}\boldsymbol T\mathrm{d}V_{x^{\prime}}\mathrm{d}V_{x}
\end{aligned}	
\end{equation}
where $\hat h(x)$ is the amount of relative elements of point $\boldsymbol x$; $\boldsymbol b$ is the body force.
\subsection{Flowchart of the proposed numerical algorithm}

\begin{figure}[h!]
	\centering
	\includegraphics[width=0.98\linewidth]{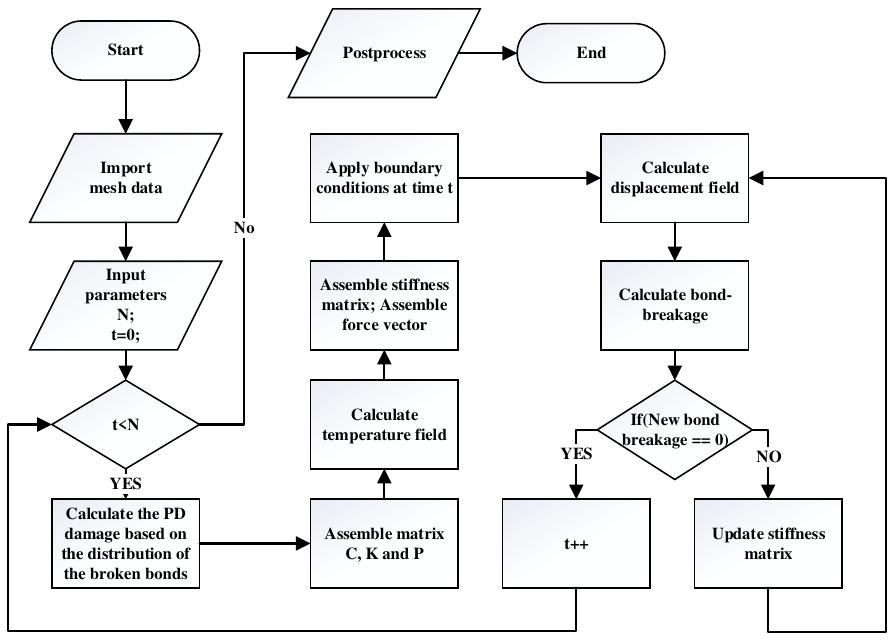}
	\caption{Flowchart of the numerical algorithm.}
	\label{fig:fig5}
\end{figure}

As depicted in the flowchart in Fig. \ref{fig:fig5}, to address thermal fracture problems based on a shared mesh system between temperature and deformation computation, the temperature field and deformation field are computed independently. At the beginning of the algorithm, the geometric model only needs to be discretized once. Subsequently, for each time step, the new definition of the PD damage is computed to characterize the degradation of the thermal conductivity. Solve the linear equation Eq. (\ref{equ:temp}) to obtain the temperature field, which is subsequently utilized to compute the equilibrium equation Eq. (\ref{equ:force}) incorporating temperature terms and obtain displacement field. The displacement field allows us to identify bond failures. Progressive accumulation of bond failures induces micro-crack nucleation. For quasi-static problems, if no new bond failure is detected, proceed to the calculation for the subsequent time step, continuing this process until the end of the computation.

\section{Numerical examples}\label{sec:sec5}
\subsection{Thermal deformation without damage}
\begin{figure}[h]
	\centering
	\includegraphics[width=0.6\linewidth]{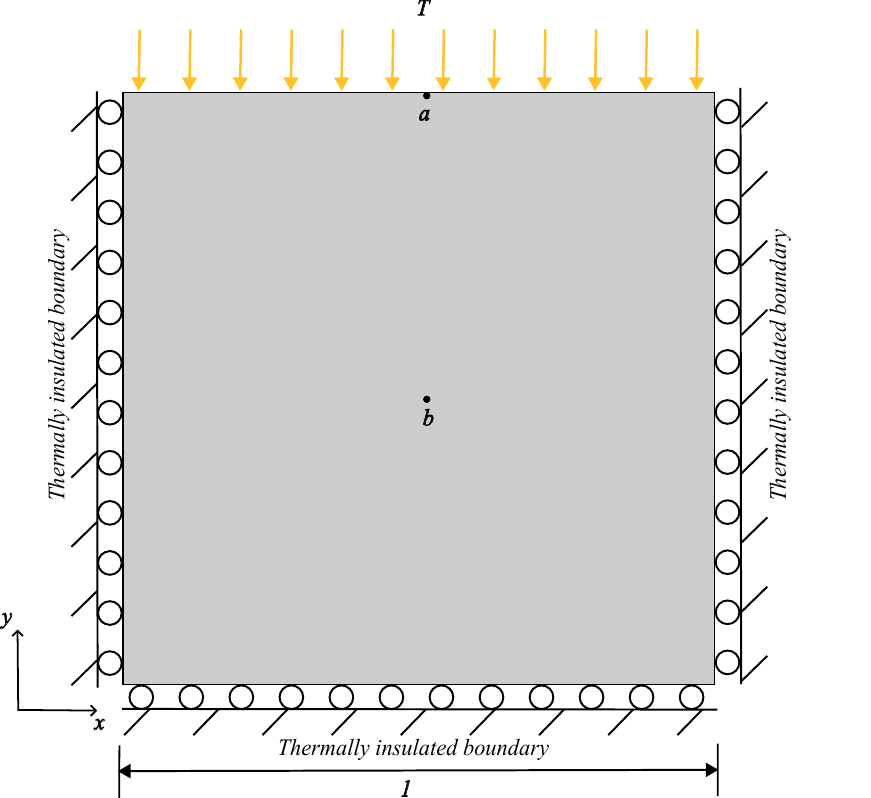}
	\caption{Sketch of the square plate (unit: m).}
	\label{fig:fig6-1}
\end{figure}

To validate the efficacy of the proposed model, we consider a scenario involving thermal deformation without damage. The condition of plane strain is adopted for this example. Timoshenko et al.  \cite{timoshenko1970a} analyzed a square plate with three edges thermally insulated and mechanically restrained against normal displacement. As illustrated in Fig. \ref{fig:fig6-1}, the top edge of the plate is subject to a Dirichlet boundary condition with a temperature $T = 1^\circ C$, while the initial temperature $T_0$ of the whole plate is $0^\circ C$. The material properties are detailed in Table  \ref{tab:tb1}. Timoshenko et al.  \cite{timoshenko1970a} and Carslaw and Jaeger \cite{Carslaw1952ConductionOH} derived the analytical solution for this problem as follows:
\begin{align}
T\left(y,t\right)=1-\frac{4}{\pi}\sum_{n=0}^{\infty}\frac{\left(-1\right)^{n}}{2n+1}\exp\left(-\frac{\left(2n+1\right)\pi^{2}kt}{4L^{2}}\right)cos\left(\frac{\left(2n+1\right)\pi y}{2L}\right)
\end{align}
and 
\begin{align}
u_{y}\left(y,t\right)=\frac{(1+\nu)}{(1-\nu)}\alpha\int_{0}^{y}T(y,t)dy
\end{align}

\begin{figure}[h!]
	\centering
	\subcaptionbox{Displacement of point $a$.}{
		\includegraphics[width=0.45\linewidth]{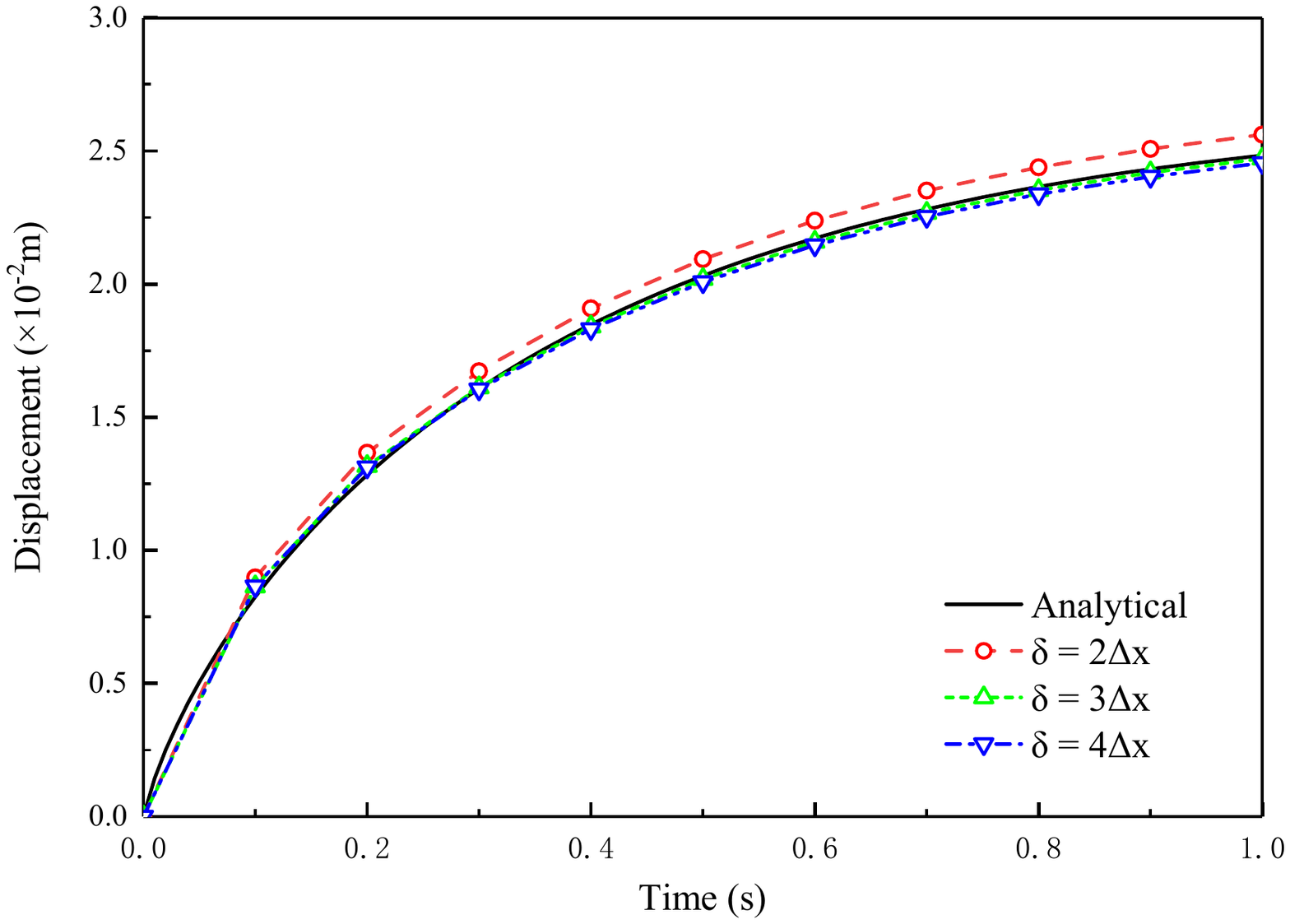}
	}
	\subcaptionbox{Displacement of point $b$.}{
		\includegraphics[width=0.45\linewidth]{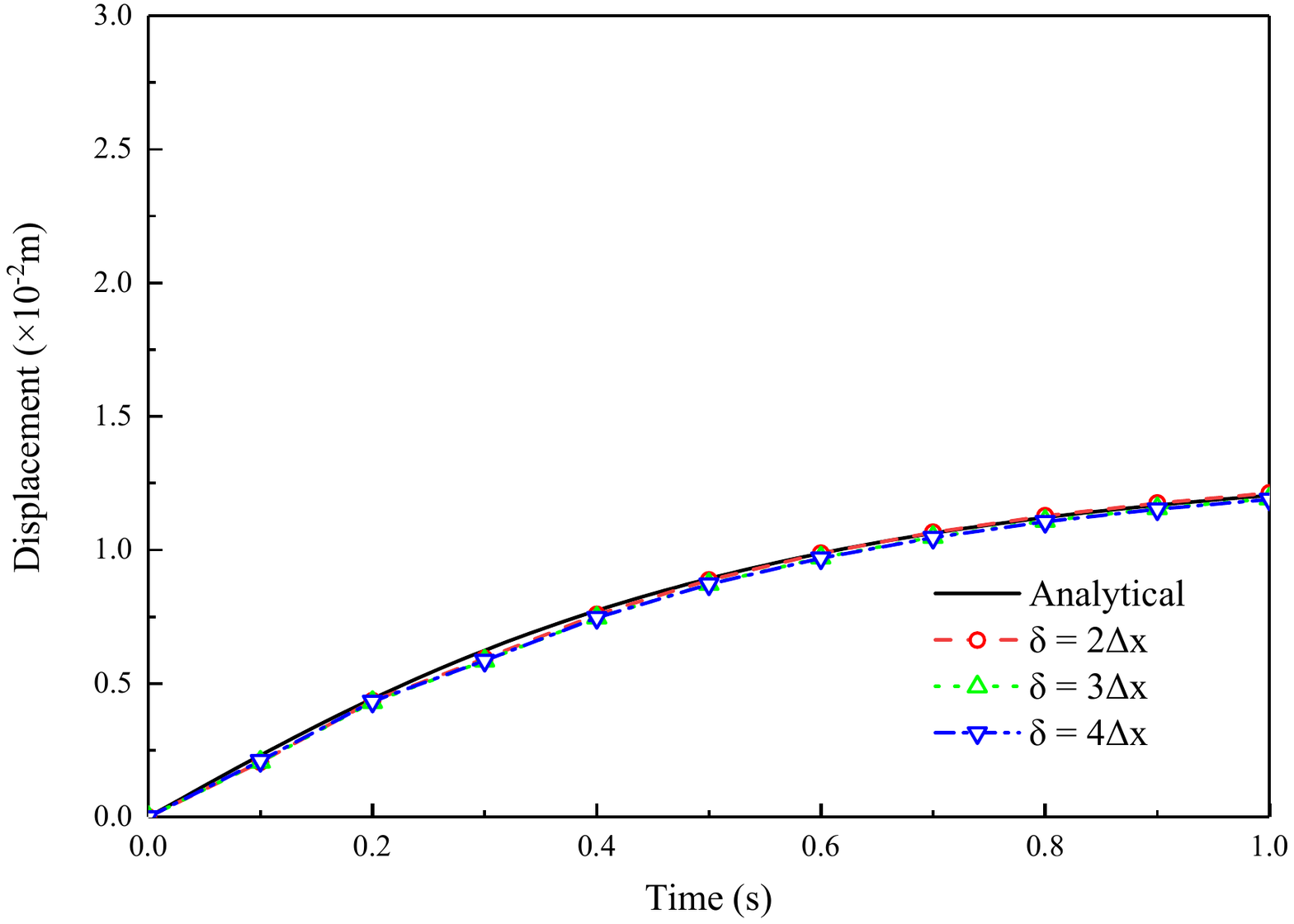}
	}
	\caption{Comparison of vertical displacement of point $a$ and $b$ between analytical and PD solutions with different horizons.}
	\label{fig:fig6-2}
\end{figure}

In this numerical example, the PD micromodulus coefficient is assumed to be an exponential function: $c^0(\|\xi\|)=\tau^0e^{-\|\xi\|/l}$, where $\tau^0$ is a constant coefficient that is calculated according to the given Poisson’s ratio and Young’s modulus, and $l$ is a characteristic length. In this example $l=\delta/3$. The model is discretized into $10,000$ uniform quadrilateral finite elements with dimensions of $10 \times 10$mm. Fig. \ref{fig:fig6-2} shows the vertical displacement of point $a$ and $b$ with different horizons, i.e., $\delta = 2\Delta x, 3\Delta x, 4\Delta x$, where $\Delta x$ is the element size. By comparing with the analytical solution, the thermo-mechanical PD model is validated. In order to improve computational efficiency and obtain accurate calculation results, $\delta = 3\Delta x$ is a suitable choice of horizon.

\begin{table}[h]
	\caption{\textbf{Material properties of the square plate.}}
	\centering
	\begin{tabular}{ccc}
		\toprule
		Parameter&Value&Unit\\
		\midrule
		Young's modulus $E$&1&$Pa$\\
		Poisson’s ratio $\nu$ &0.25&–\\
		Density $\rho$&0.0&$kg/m^3$\\
		Thermal conductivity $k$&1.0&$J/(s\cdot m\cdot K)$\\
		Specific heat capacity $c$&1.0&$J/(kg \cdot K)$\\
		Thermal expansion coefficient $\alpha$&0.016&$1/K$\\
		\bottomrule
	\end{tabular}
	\label{tab:tb1}
\end{table}

\subsection{Heat flow in a plate with two thermal insulation cracks}
\begin{figure}[h!]
	\centering
	\includegraphics[width=0.6\linewidth]{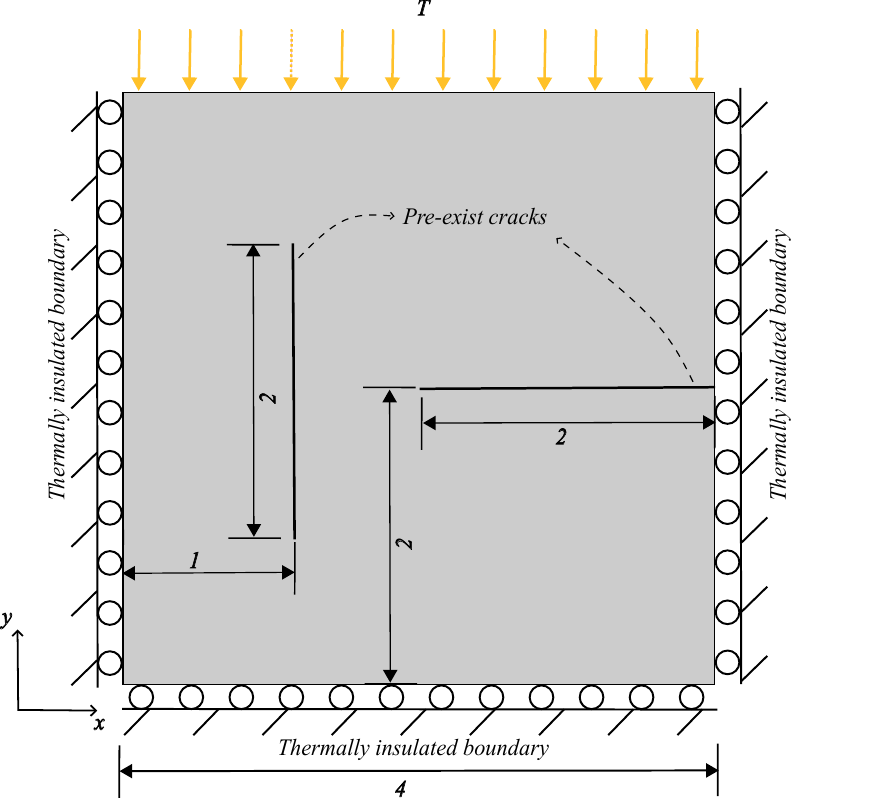}
	\caption{Sketch of the plate with thermal insulation cracks (unit: m).}
	\label{fig:fig7}
\end{figure}

To validate the proposed new definition of the PD damage, heat flow in a plate containing thermal insulation cracks is considered. The geometry configuration and boundary conditions are shown in Fig. \ref{fig:fig7}. There are two pre-exist thermal insulation cracks, one oriented horizontally and the other vertically. The top edge of the plate is subjected to a Dirichlet boundary condition with a temperature of  $T = 100^\circ C$. The initial temperature $T_0$ of the whole plate is $0^\circ C$. The material properties are shown in Table \ref{tab:tb1}. The model is discretized into 10,000 uniform quadrilateral finite elements with dimensions of $40 \times 40$mm. The horizon is $\delta = 3\Delta x$.

The classical thermal insulation cracks are modeled through thermal conductivity degradation as:
\begin{align}\label{equ:sun1}
	k=(1-\overline{d})^2 k_0
\end{align}
where $\overline{d}$ is defined as:
\begin{align}\label{equ:sun2}
	\overline{d}=\begin{cases}\quad0&\text{if }d\leqslant c_1\\\frac{d-c_1}{c_2-c_1}&\text{if }c_1<d\leqslant c_2\\\quad1&\text{if }d>c_2\end{cases}
\end{align}
here, $c_1$ and $c_2$ are two threshold values. Although above model can capture the impact of thermal insulation cracks on the temperature field, it assumes that the influence of cracks on thermal conductivity is consistent in all directions, which may not always be the case in practice.

\begin{figure}[h!]
	\centering
	\subcaptionbox{Heat flux calculated by the classical model for thermal insulation cracks.}{
		\includegraphics[height=0.44\linewidth]{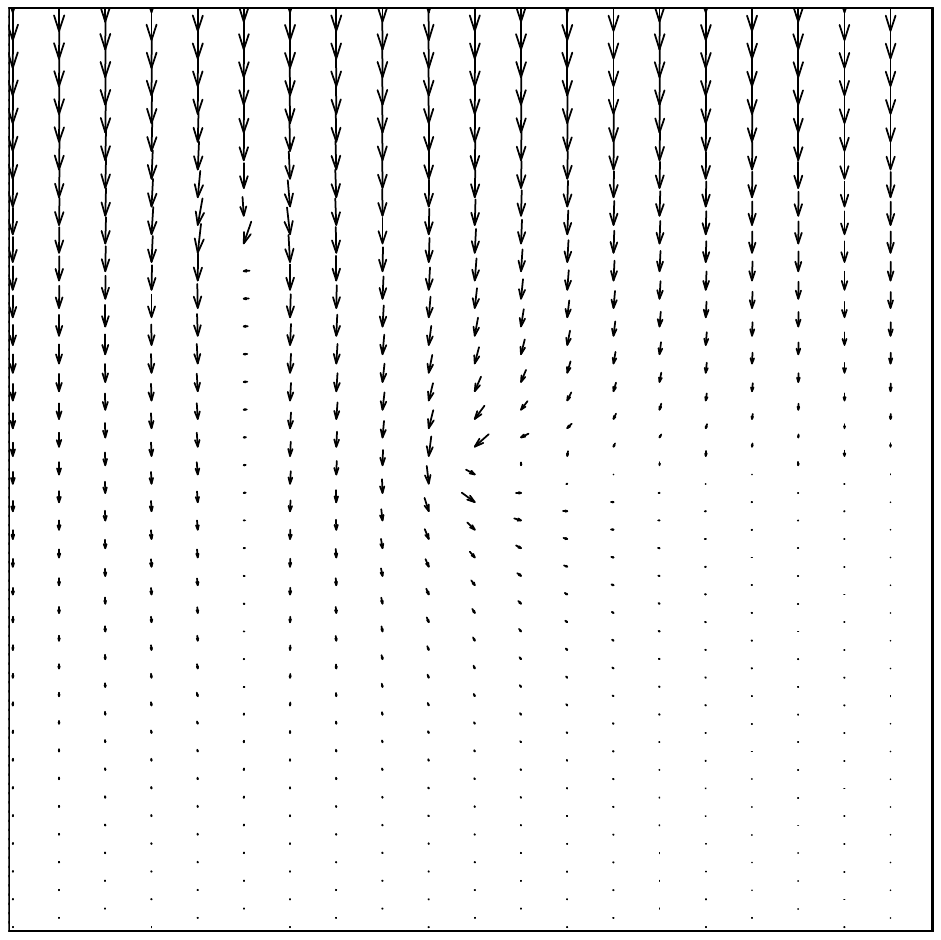}
	}
	\subcaptionbox{Heat flux calculated by the proposed model for thermal insulation cracks.}{
		\includegraphics[height=0.44\linewidth]{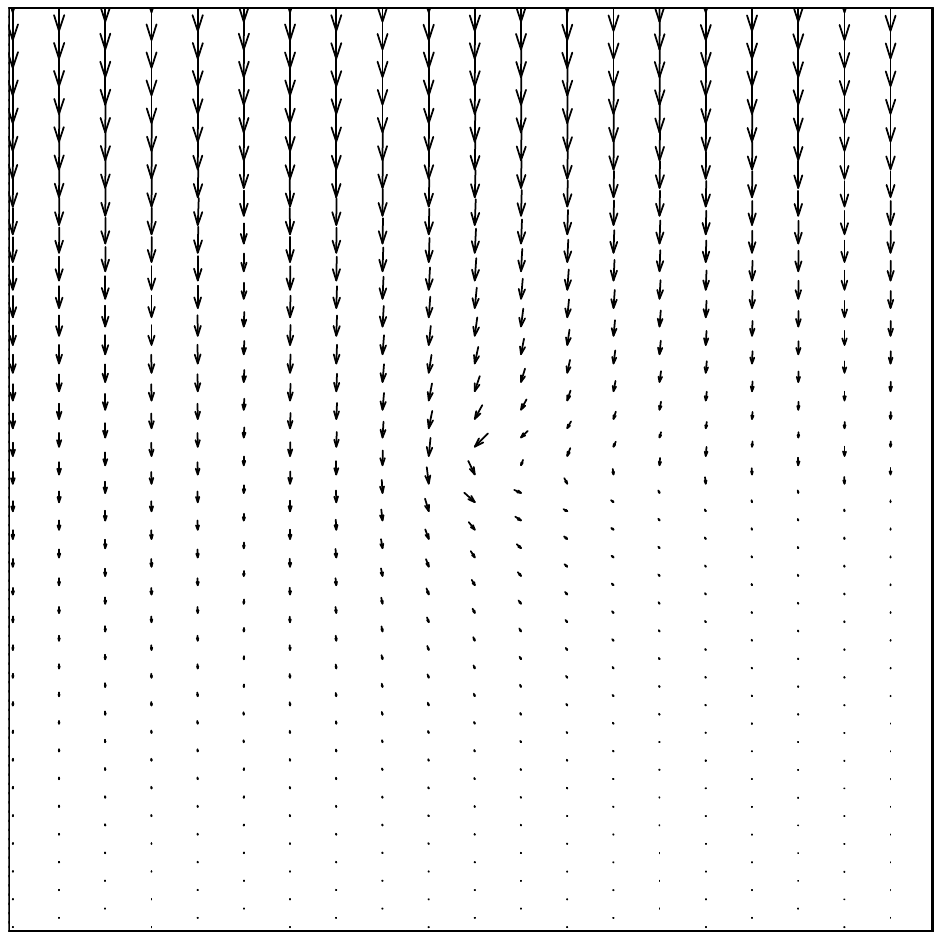}
	}
	\caption{Comparison of heat flux at $t=1 $s between two models.}
	\label{fig:fig7-2}
\end{figure}

\begin{figure}[h!]
	\centering
	\subcaptionbox{$k_x$ calculated by the classical model for thermal insulation cracks.}{
		\includegraphics[height=0.44\linewidth]{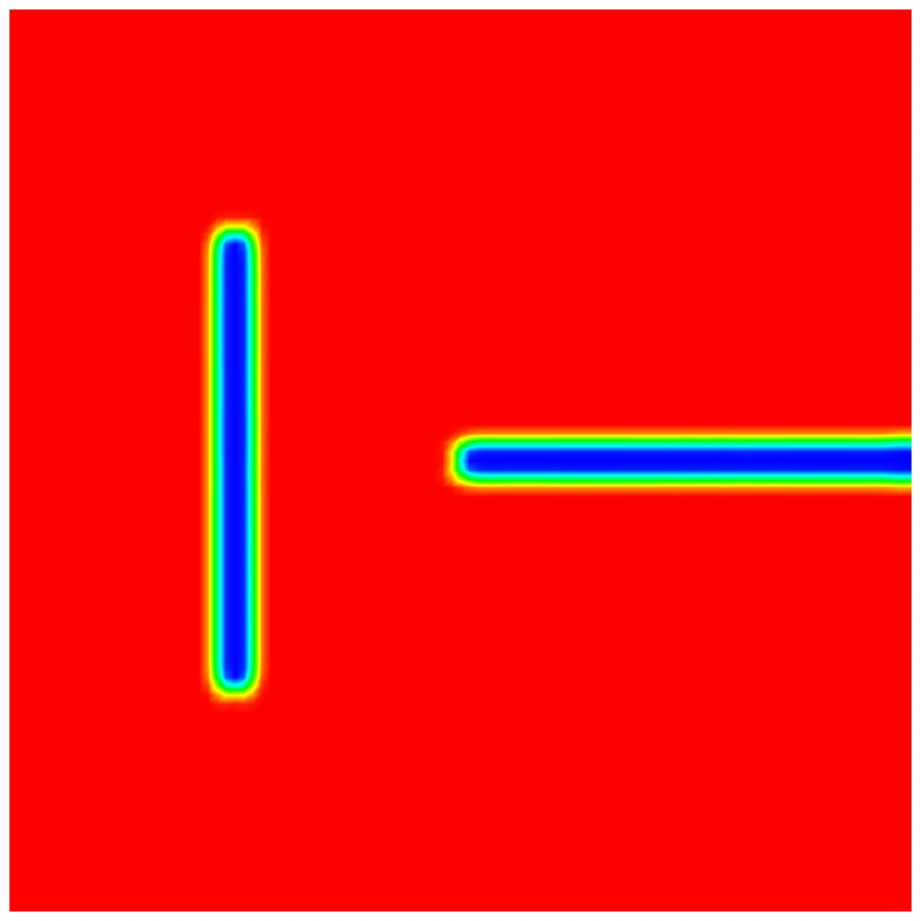}
	}
	\subcaptionbox{$k_x$ calculated by the proposed model for thermal insulation cracks.}{
		\includegraphics[height=0.44\linewidth]{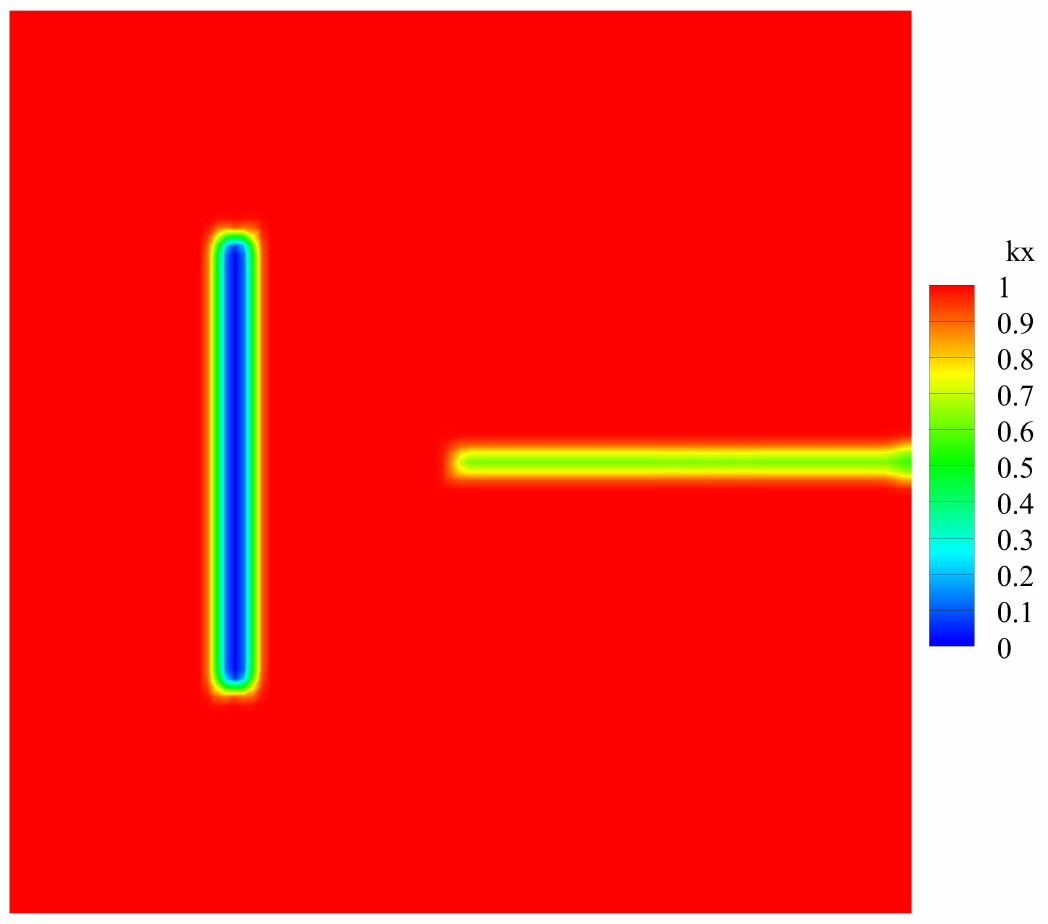}
	}
	\subcaptionbox{$k_y$ calculated by the classical model for thermal insulation cracks.}{
		\includegraphics[height=0.44\linewidth]{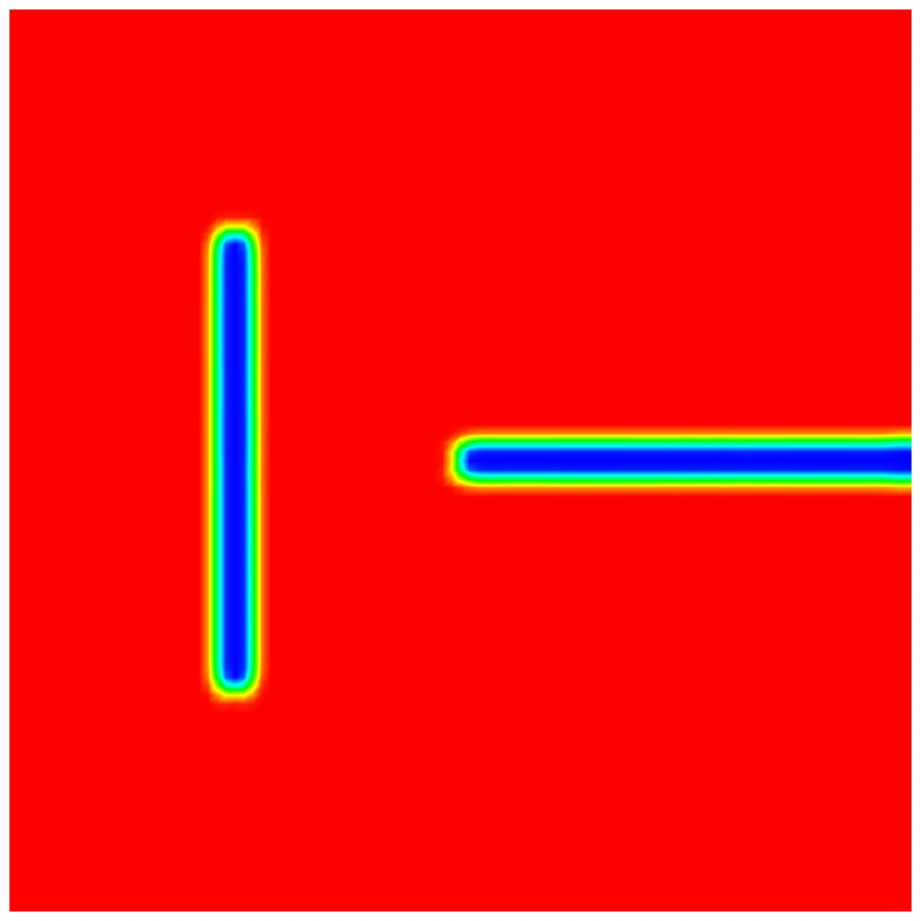}
	}
	\subcaptionbox{$k_y$ calculated by the proposed model for thermal insulation cracks.}{
		\includegraphics[height=0.44\linewidth]{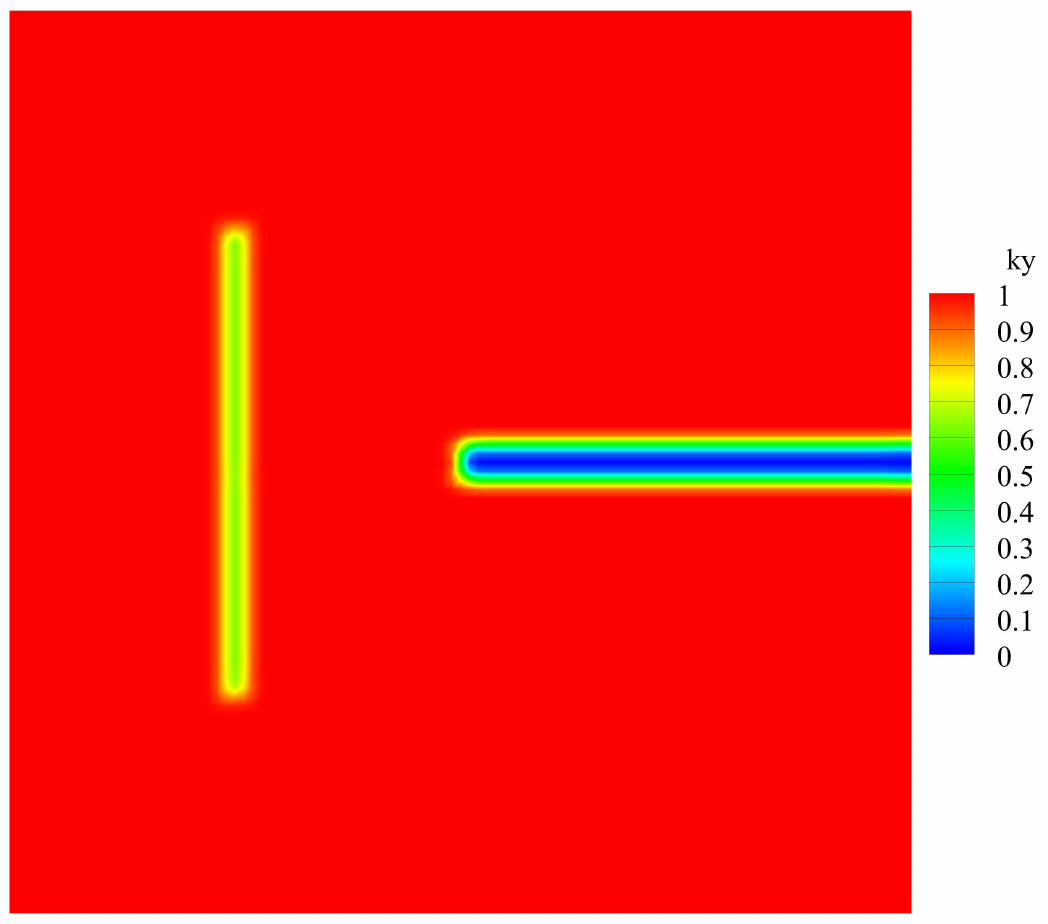}
	}
	\caption{Comparison of thermal conductivity between two models.}
	\label{fig:fig7-3}
\end{figure}

\begin{figure}[h!]
	\centering
	\subcaptionbox{Temperature field contour calculated by the classical model for thermal insulation cracks.}{
		\includegraphics[height=0.44\linewidth]{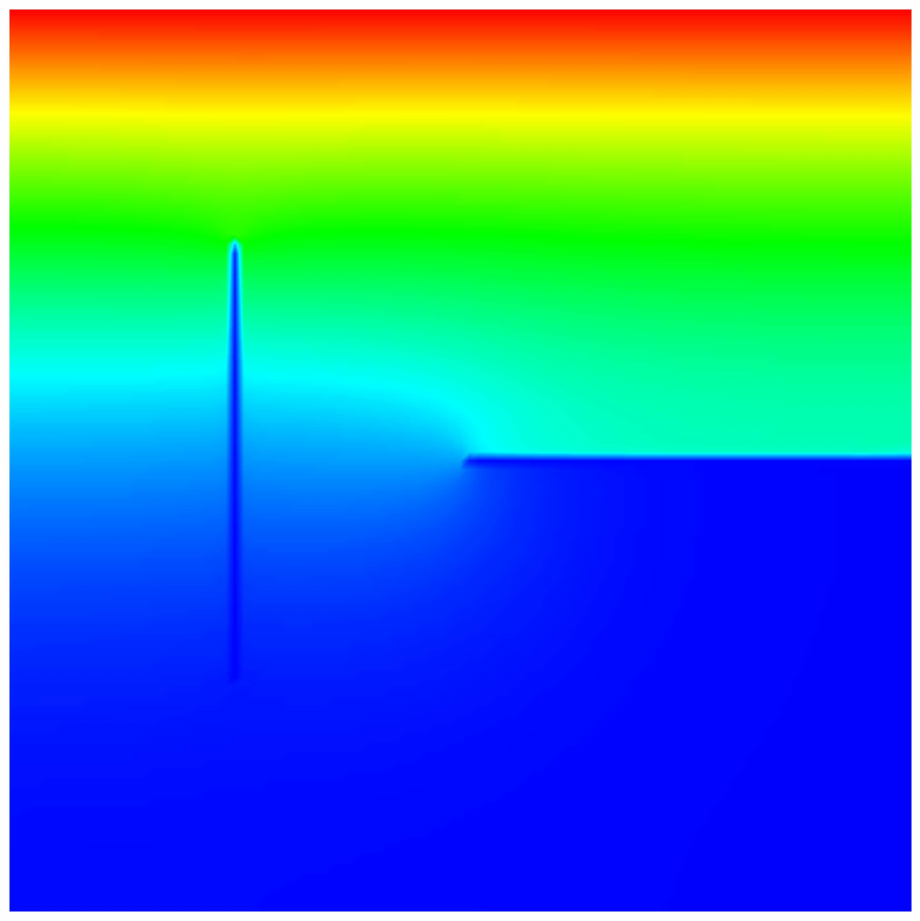}
	}
	\subcaptionbox{Temperature field contour calculated by the proposed model for thermal insulation cracks.}{
		\includegraphics[height=0.44\linewidth]{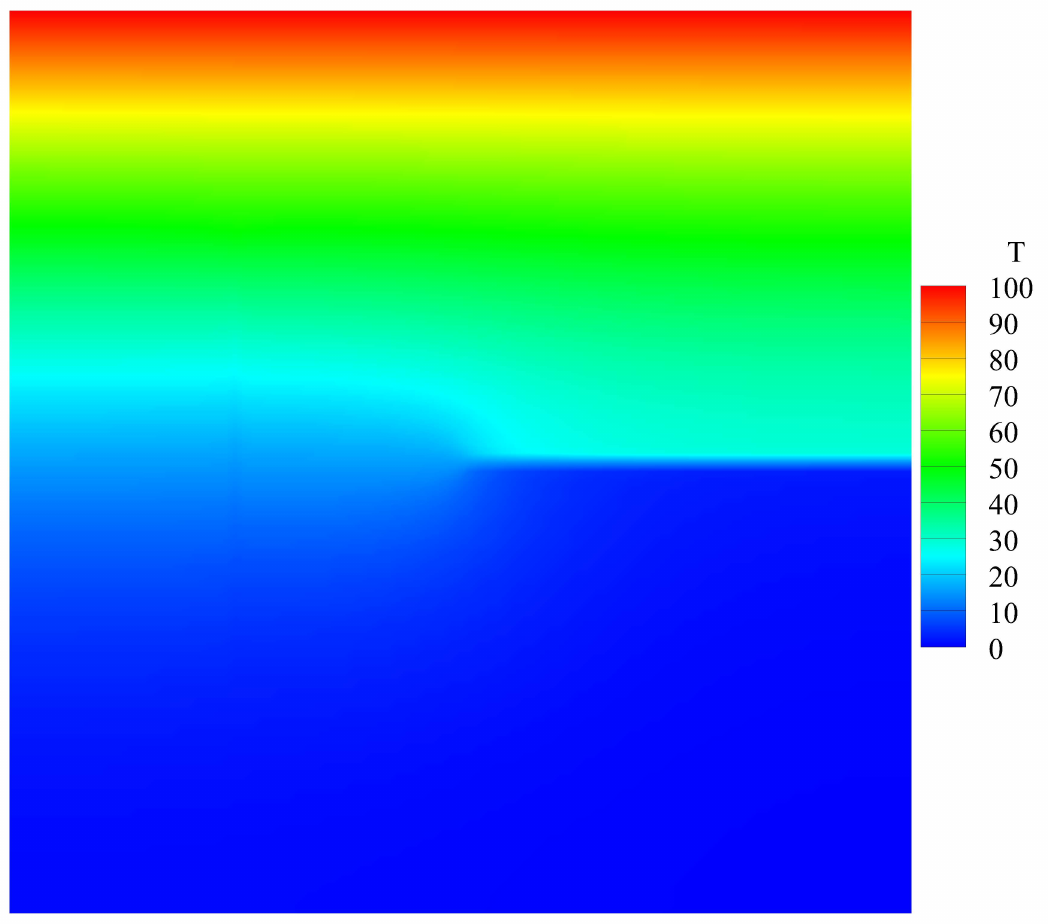}
	}
	\caption{Comparison of temperature field contours between two models (unit: $^\circ C$).}
	\label{fig:fig7-5}
\end{figure}

The calculation results by Eqs. (\ref{equ:sun1}) and (\ref{equ:sun2}) and the proposed model by Eq. (\ref{equ:k}) for thermal insulation cracks are compared. Fig. \ref{fig:fig7-2} shows the heat flux calculation results between the classical and the proposed model for thermal insulation cracks. Fig. \ref{fig:fig7-3} shows the comparison of thermal conductivity $k_x$ and $k_y$ between two models. Notably, the classical model calculates thermal conductivity based on a scalar damage value, resulting in equal thermal conductivity in both directions. In contrast, the new definition of the PD damage captures directional differences in thermal conductivity. Specifically, the pre-exist crack in the y-direction significantly affects the thermal conductivity in the x-direction but has minimal impact on the thermal conductivity in the y-direction. Conversely, the effect of cracks in x-direction on thermal conductivity is opposite. Fig.  \ref{fig:fig7-5} compares the temperature field contours between two models, further demonstrating the necessity and advantages of the proposed model.

\subsection{Thermal fracture of a cruciform plate with a corner crack}
\begin{figure}[h!]
	\centering
	\subcaptionbox{Condition 1.}{
		\includegraphics[width=0.45\linewidth]{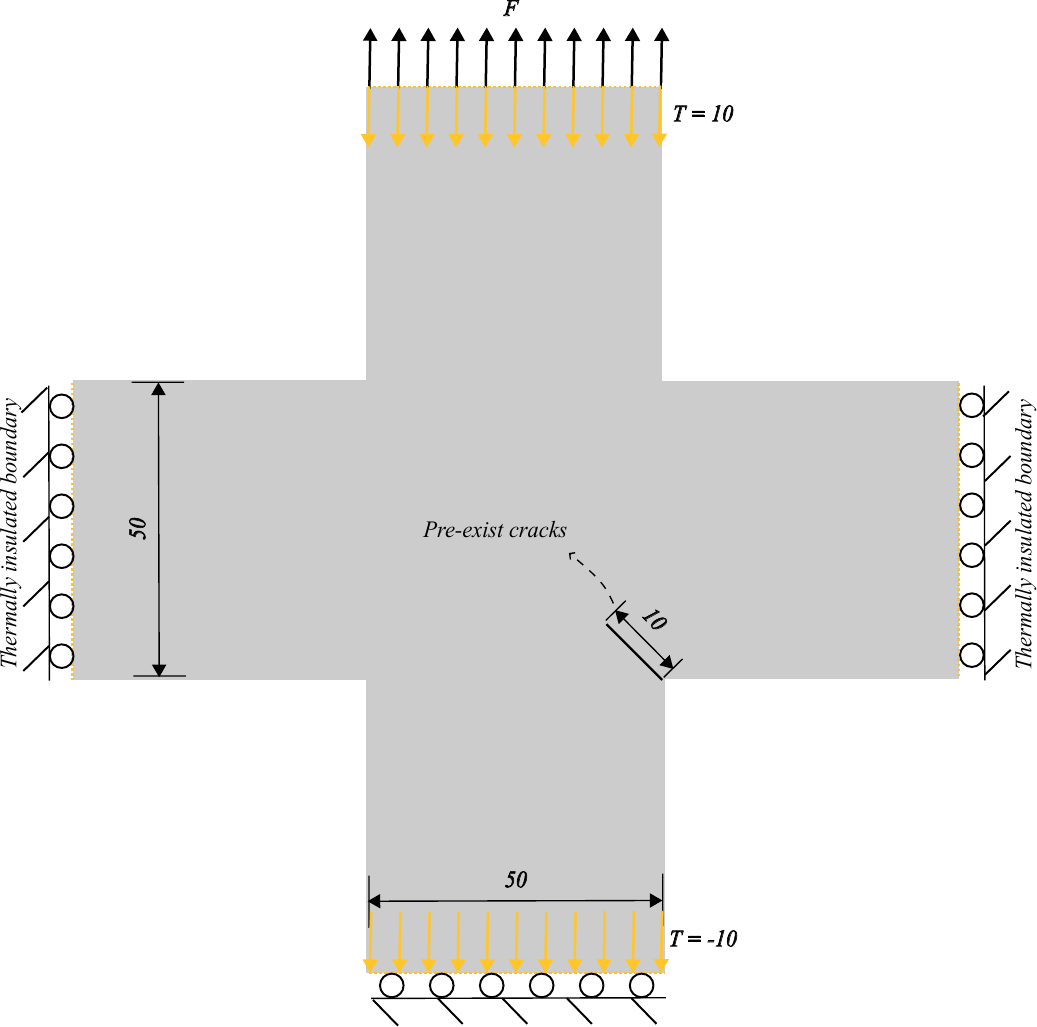}
	}
	\subcaptionbox{Condition 2.}{
		\includegraphics[width=0.45\linewidth]{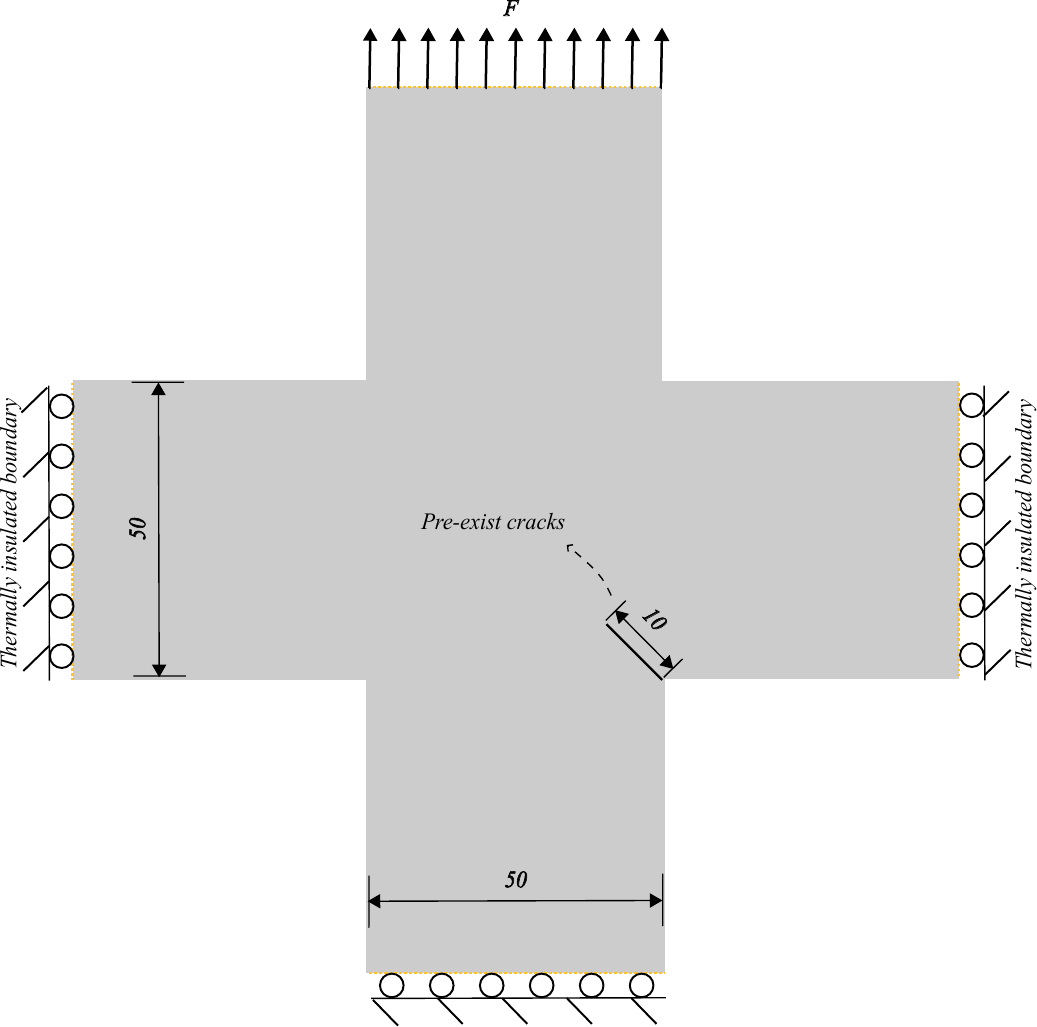}
	}
	\subcaptionbox{Condition 3.}{
		\includegraphics[width=0.45\linewidth]{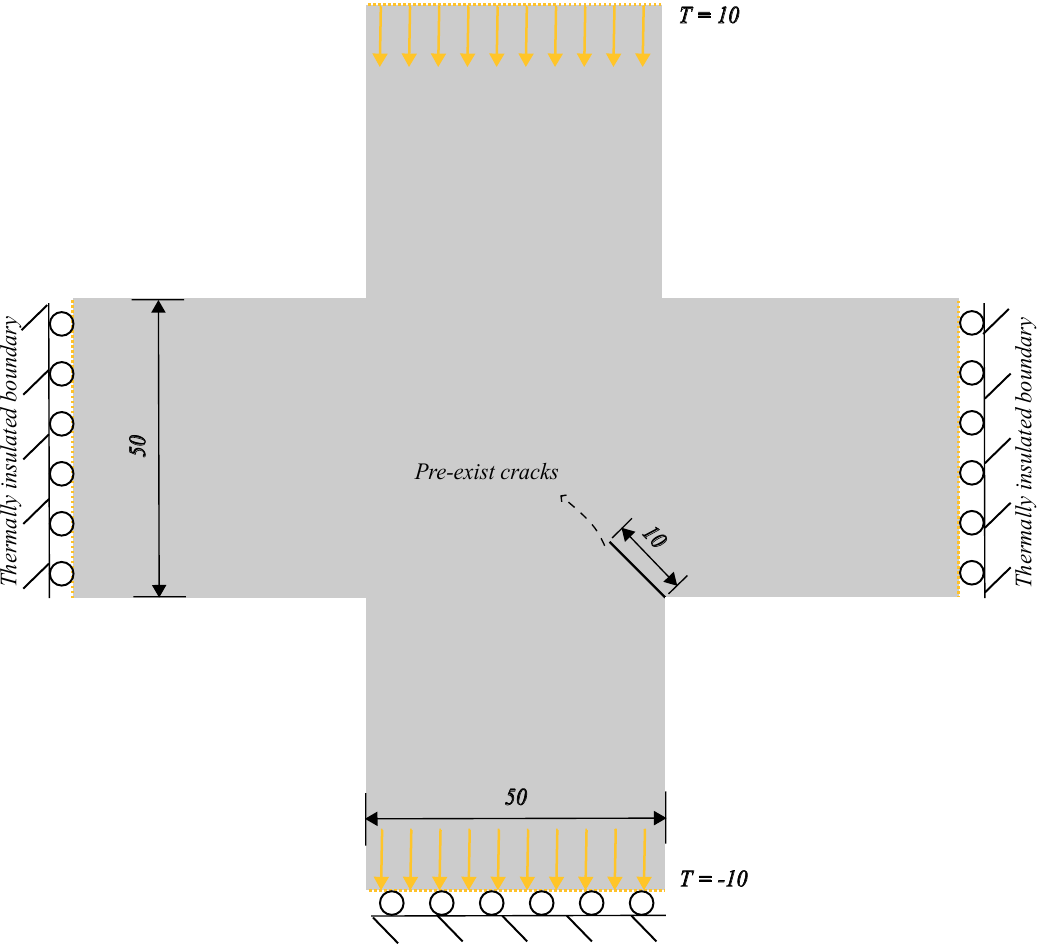}
	}
	\caption{Sketch of the cruciform plate with a corner crack (unit: mm).}
	\label{fig:fig8}
\end{figure}

\begin{figure}[h!]
	\centering
	\subcaptionbox{Temperature field contour calculated by classical thermal insulation cracks model.}{
		\includegraphics[height=0.44\linewidth]{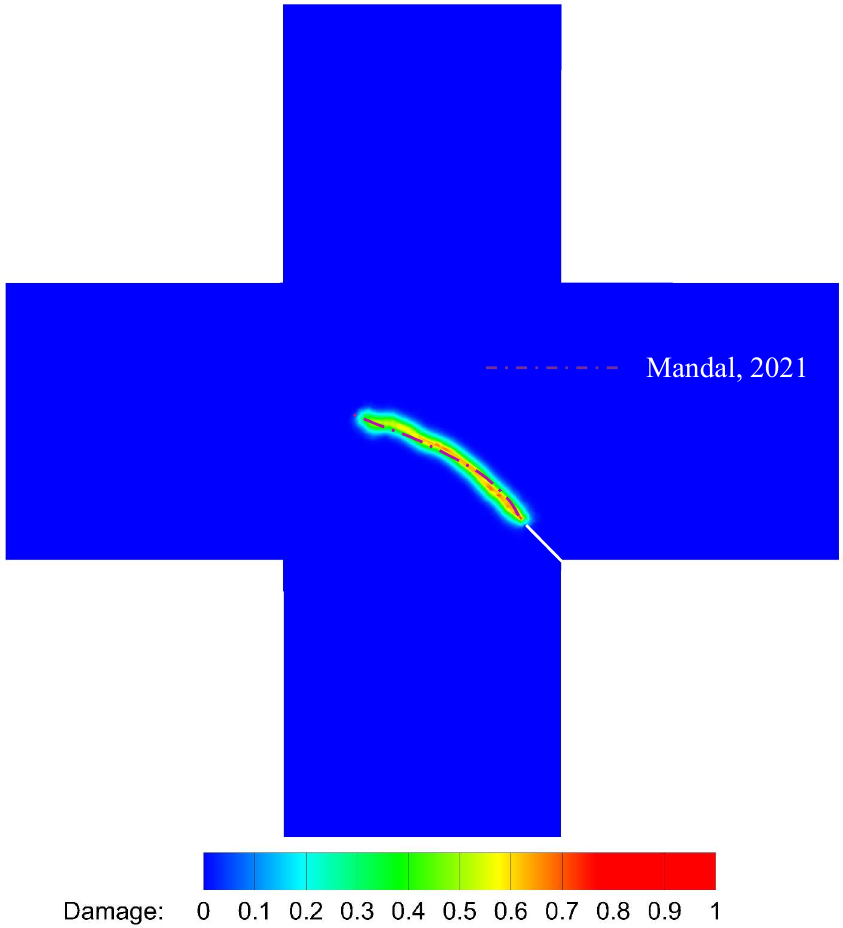}
	}
	\subcaptionbox{Temperature field contour calculated by the proposed model.}{
		\includegraphics[height=0.44\linewidth]{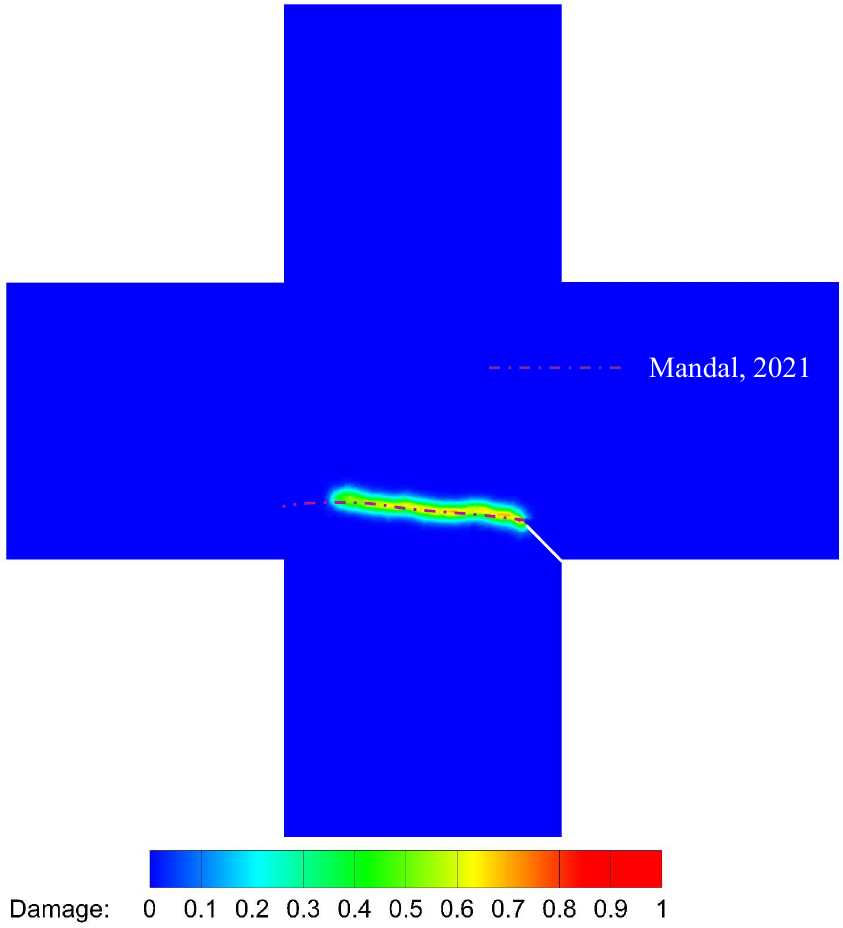}
	}
	\caption{Comparison of temperature field contour between classical thermal insulation cracks model and the proposed model.}
	\label{fig:crackplate}
\end{figure}

\begin{figure}[h!]
	\centering
	\includegraphics[width=0.9\linewidth]{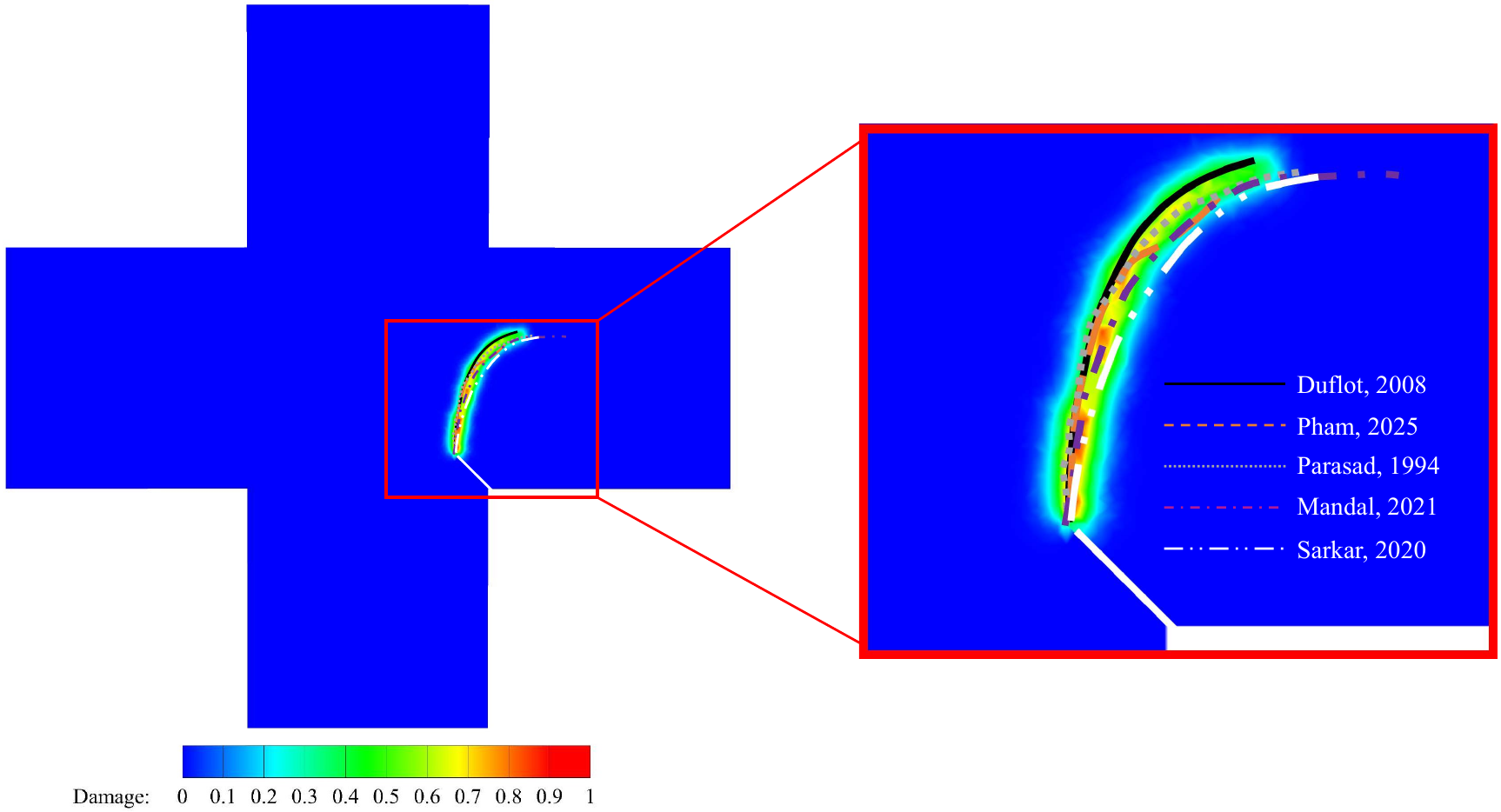}
	\caption{Comparison of crack paths between Mandal et al. \cite{mandal_fracture_2021}, Duflot et al. \cite{duflot_extended_2008}, Pham et al. \cite{pham_adaptive_2025}, Prasad et al. \cite{prasad_incremental_1994}, Sarkar et al. \cite{sarkar2020} and the present model (represented by colored contour).}
	\label{fig:crackplate2}
\end{figure}

\begin{figure}[h!]
	\centering
	\subcaptionbox{Thermal conductivity $k_x$.}{
		\includegraphics[width=0.3\linewidth]{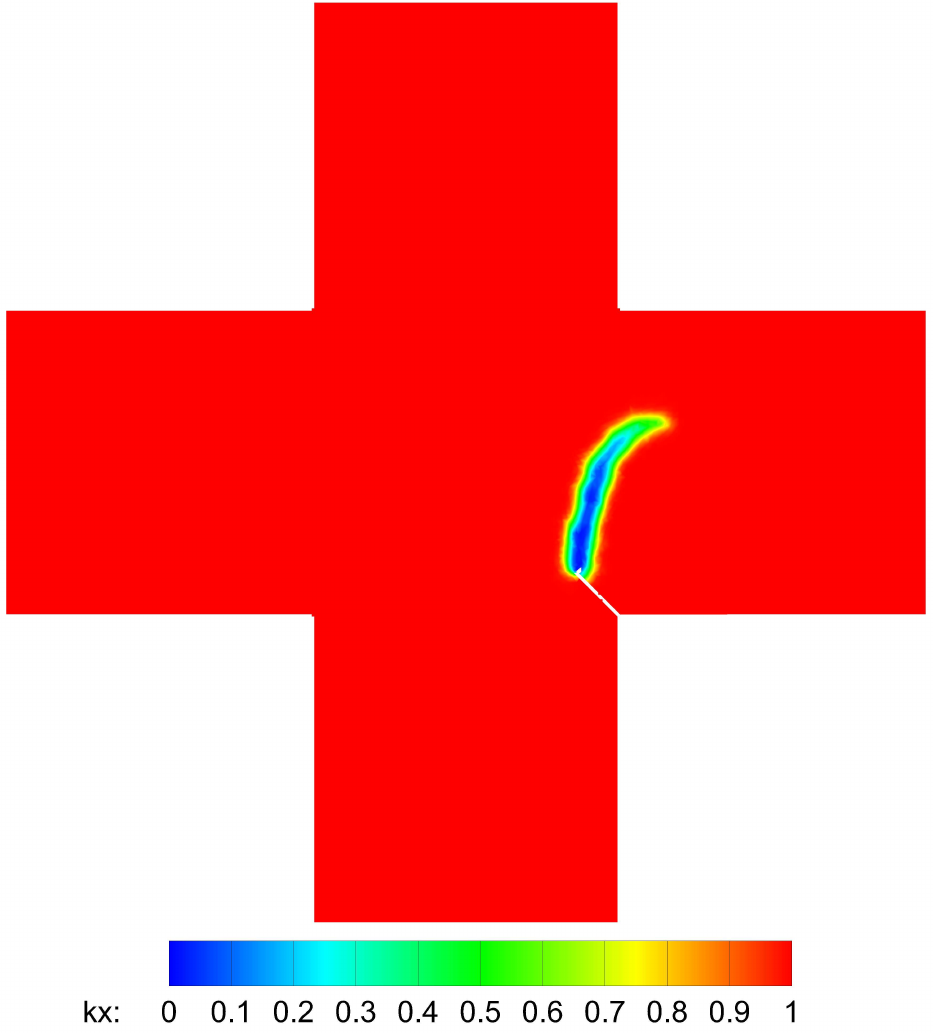}
	}
	\subcaptionbox{Thermal conductivity $k_y$.}{
		\includegraphics[width=0.3\linewidth]{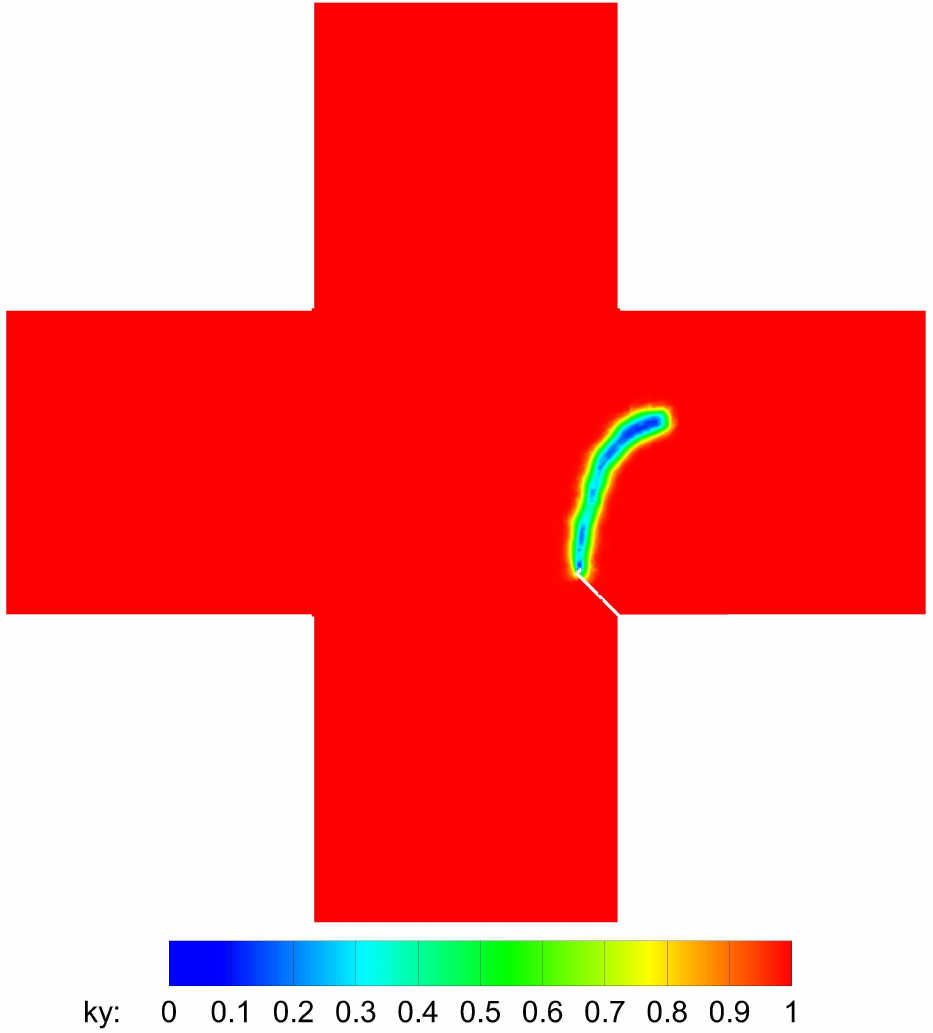}
	}
	\subcaptionbox{Temperature field contour.}{
		\includegraphics[width=0.3\linewidth]{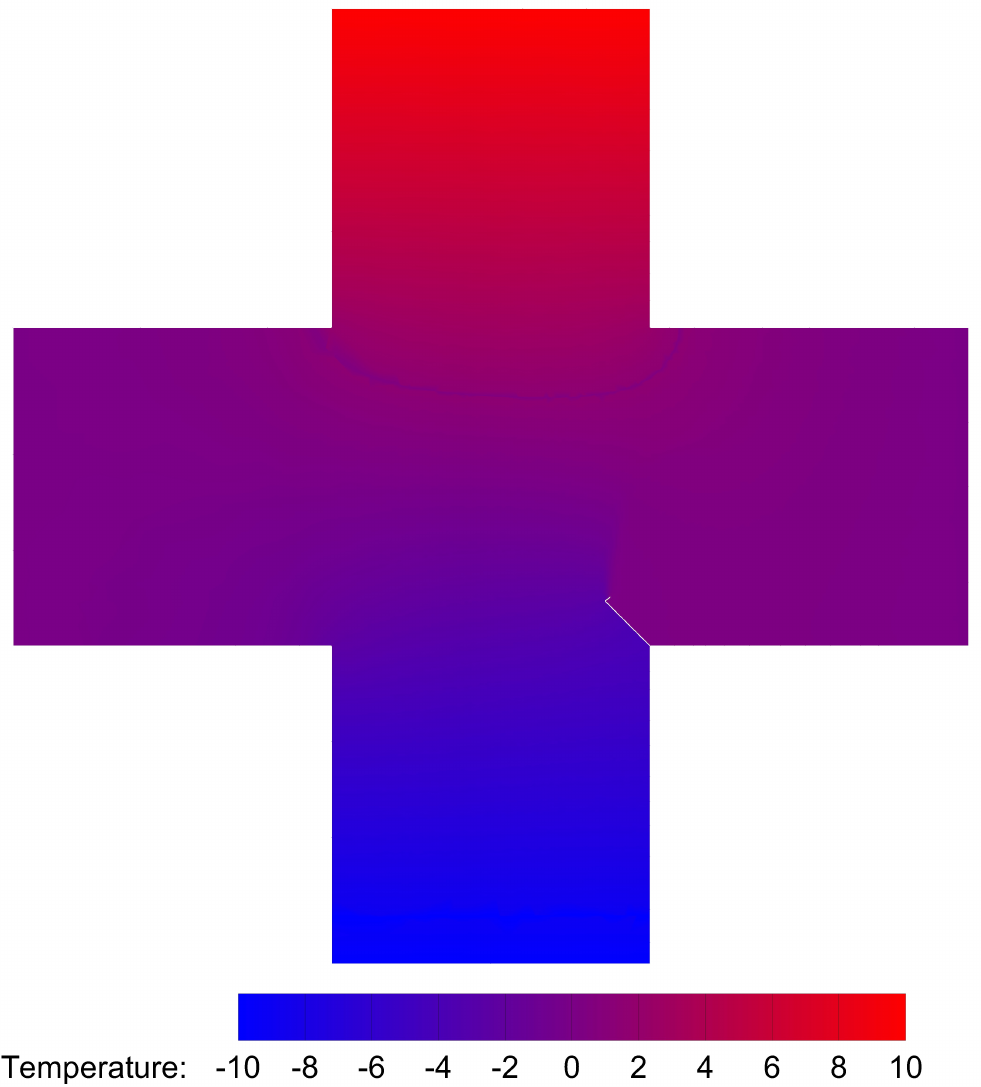}
	}
	\caption{Calculation results of thermal conductivity and temperature field of the proposed model.}
	\label{fig:crackplate3}
\end{figure}

This example is a quasi-static crack propagation problem under thermo-mechanical conditions. Fig. \ref{fig:fig8} depicts a cruciform plate with a corner crack, where the crack length is 10mm and forms a $45^\circ$ angle with the vertical axis. The numerical computations are conducted under the assumption of plane stress. We consider crack propagation paths under three distinct mechanical and thermal boundary conditions. In all three conditions, the initial temperature is set to $0^\circ C$, and the three edges of the plate are mechanically restrained against normal displacement. The material properties are shown in Table \ref{tab:tb3}, while the boundary conditions for Fig. \ref{fig:fig8} are detailed in Table \ref{tab:tb4}. To accelerate the calculations, we introduce PD domain only in the vicinity of the corner crack and employ the CCM model in the remaining domain. The PD domain is discretized into quadrilateral finite element meshes with dimensions of $1 \times 1$mm, while the CCM domain is discretized into finite element meshes with dimensions of $5 \times 5$mm. The PD horizon is set to $\delta = 3\Delta x$. Fig. \ref{fig:crackplate} displays the final crack paths under conditions 1 and 2. These crack paths are compared with those predicted by PFM proposed by Mandal et al.  \cite{mandal_fracture_2021}, showing good agreement between the two models. For condition 3, as illustrated in Fig. \ref{fig:crackplate2}, we compare the crack paths obtained from various methods: PFM by Mandal et al. \cite{mandal_fracture_2021}, XFEM by Duflot et al. \cite{duflot_extended_2008},  adaptive mesh refinement (AMR) method by Pham et al. \cite{pham_adaptive_2025}, boundary element method (BEM) by Prasad et al. \cite{prasad_incremental_1994}, gradient-enhanced damage method by Sarkar et al. \cite{sarkar2020}, and the present method (represented by colored contours). The new definition of the PD damage, which can distinguish damage in two directions, enables a more accurate description of the effect of thermal insulation crack on thermal conductivity. Fig. \ref{fig:crackplate3}(a) and \ref{fig:crackplate3}(b) show the degradation of thermal conductivity $k_x$ and $k_y$ due to the thermal insulation crack. Fig. \ref{fig:crackplate3}(c) shows the temperature field.

\begin{table}[h!]
	\caption{\textbf{Material properties of the cruciform plate}}
	\centering
	\begin{tabular}{ccc}
		\toprule
		Parameter&Value&Unit\\
		\midrule
		Young's modulus $E$&218.4e3&$Pa$\\
		Poisson’s ratio $\nu$ &0.33&–\\
		Thermal conductivity $k$&1.0&$J/(s\cdot m\cdot K)$\\
		Specific heat capacity $c$&1.0&$J/(kg \cdot K)$\\
		Thermal expansion coefficient $\alpha$&6.0e-4&$1/K$\\
		Fracture energy $G$&2.0e-4&$N/m$\\
		\bottomrule
	\end{tabular}
	\label{tab:tb2}
\end{table}

\begin{table}[h!]
	\caption{\textbf{Three boundary conditions of the cruciform plate}}
	\centering
	\begin{tabular}{cccc}
		\toprule
		&\multicolumn{2}{c}{Temperature ($^\circ C$)}&\\
		{BC}&upper&bottom&{Displacement of the top edge (mm)}\\
		\midrule
		1&10&-10&5e-4\\
		2&0&0&5e-4\\
		3&10&-10&-\\
		\bottomrule
	\end{tabular}
	\label{tab:tb3}
\end{table}


\subsection{Thermal shock fractures in ceramics}
\begin{figure}[h!]
	\centering
	\includegraphics[width=0.6\linewidth]{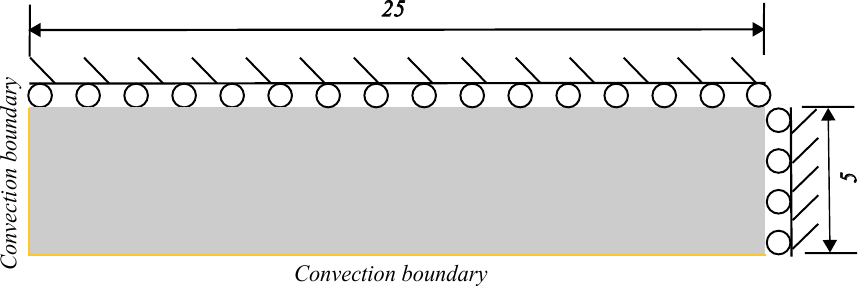}
	\caption{Sketch of thin ceramics plate (unit: mm).}
	\label{fig:fig9}
\end{figure}

\begin{figure}[h!]
	\centering
	\subcaptionbox{$T_0=300^\circ C$, $d_x$.}{
		\includegraphics[width=0.3\linewidth]{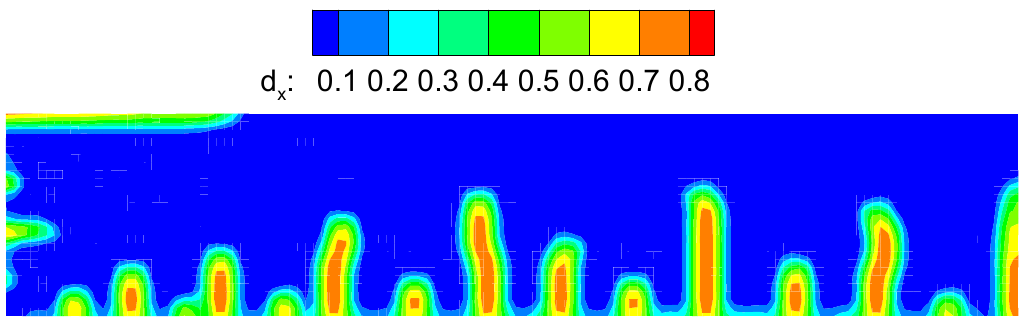}
	}
	\subcaptionbox{$T_0=300^\circ C$, $d_y$.}{
		\includegraphics[width=0.3\linewidth]{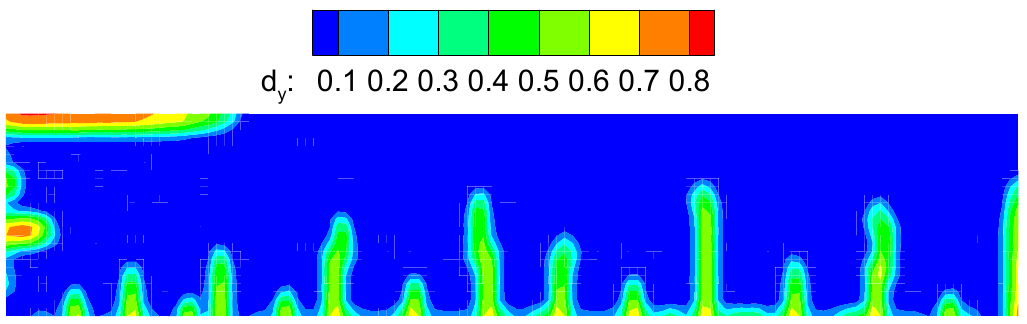}
	}
	\subcaptionbox{$T_0=300^\circ C$.}{
		\includegraphics[width=0.3\linewidth]{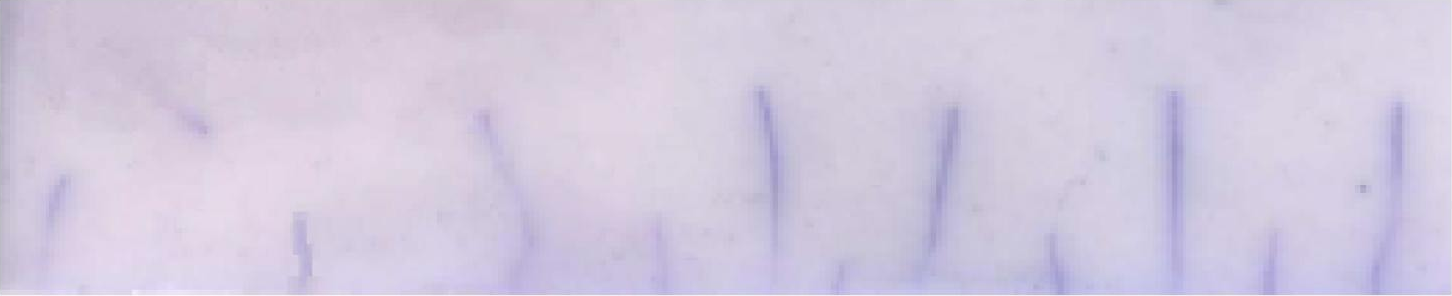}
	}
	\subcaptionbox{$T_0=350^\circ C$, $d_x$.}{
	\includegraphics[width=0.3\linewidth]{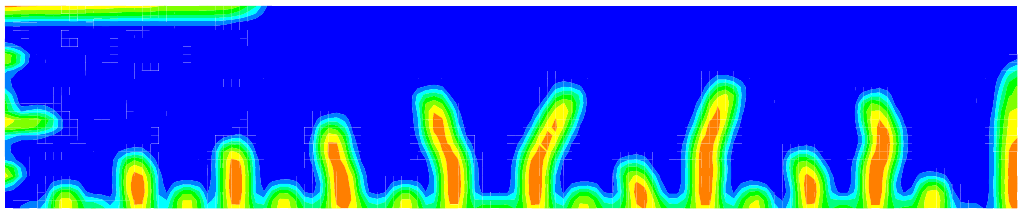}
}
	\subcaptionbox{$T_0=350^\circ C$, $d_y$.}{
	\includegraphics[width=0.3\linewidth]{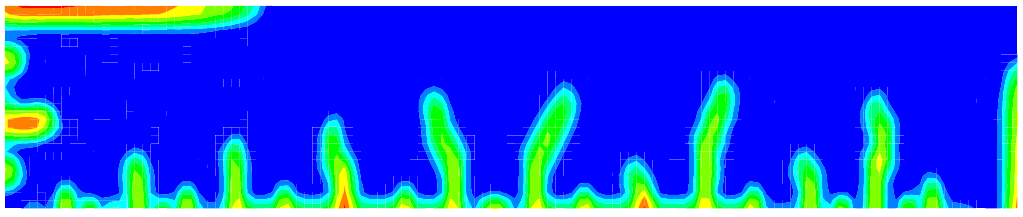}
}
	\subcaptionbox{$T_0=350^\circ C$.}{
	\includegraphics[width=0.3\linewidth]{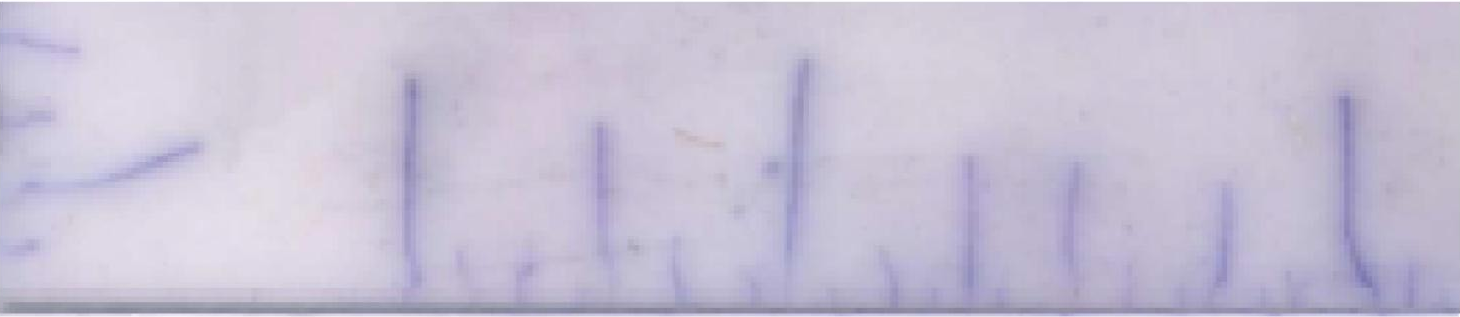}
}
	\subcaptionbox{$T_0=400^\circ C$, $d_x$.}{
	\includegraphics[width=0.3\linewidth]{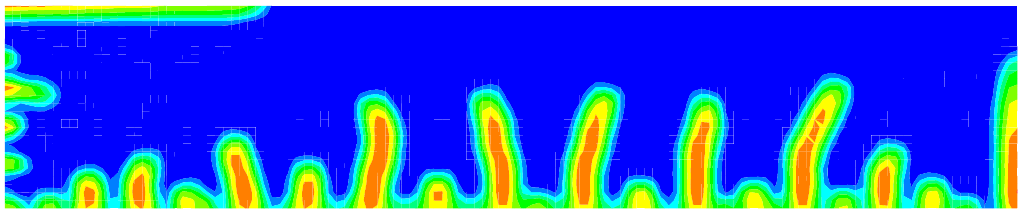}
}
	\subcaptionbox{$T_0=400^\circ C$, $d_y$.}{
	\includegraphics[width=0.3\linewidth]{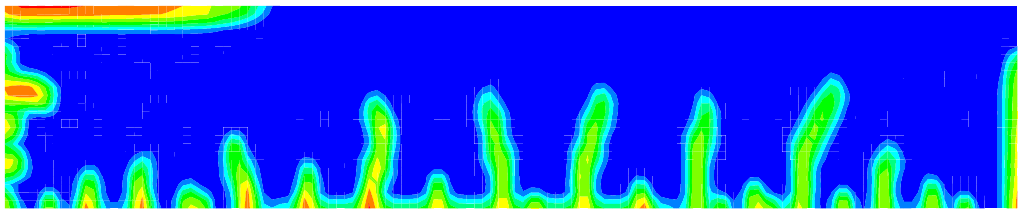}
}
	\subcaptionbox{$T_0=400^\circ C$.}{
	\includegraphics[width=0.3\linewidth]{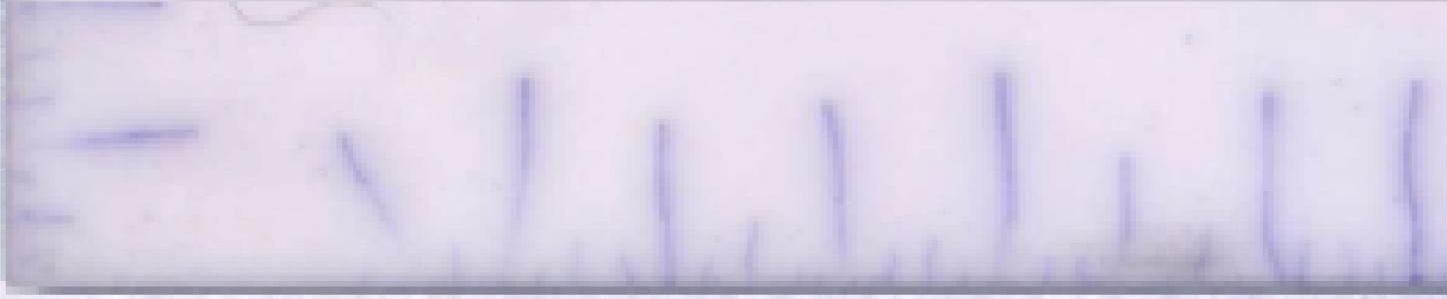}
}
	\subcaptionbox{$T_0=500^\circ C$, $d_x$.}{
	\includegraphics[width=0.3\linewidth]{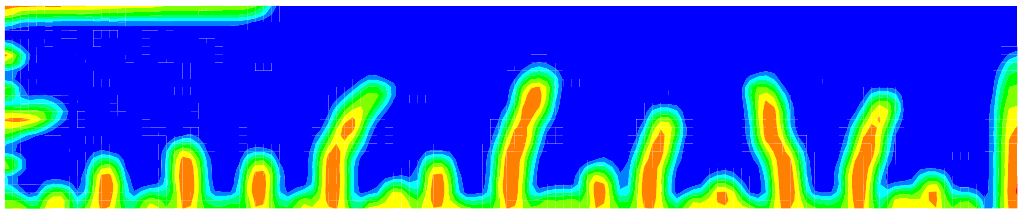}
}
	\subcaptionbox{$T_0=500^\circ C$, $d_y$.}{
	\includegraphics[width=0.3\linewidth]{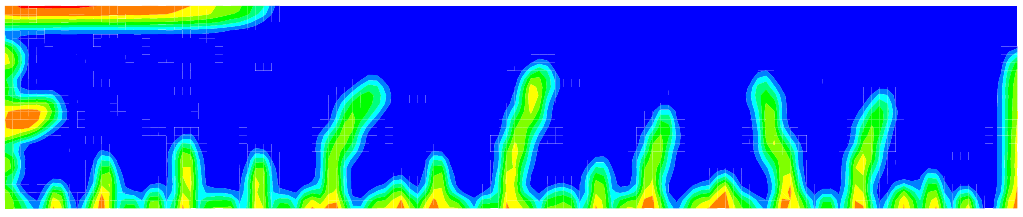}
}
	\subcaptionbox{$T_0=500^\circ C$.}{
	\includegraphics[width=0.3\linewidth]{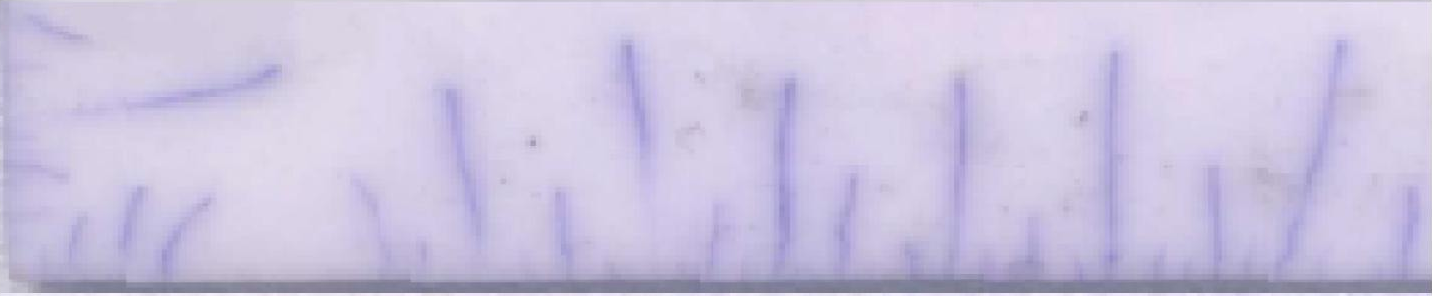}
}
	\subcaptionbox{$T_0=600^\circ C$, $d_x$.}{
	\includegraphics[width=0.3\linewidth]{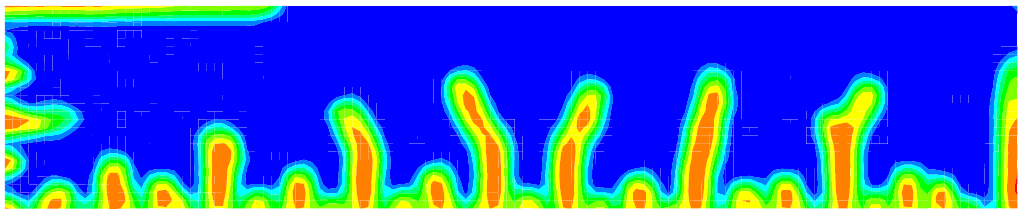}
}
	\subcaptionbox{$T_0=600^\circ C$, $d_y$.}{
	\includegraphics[width=0.3\linewidth]{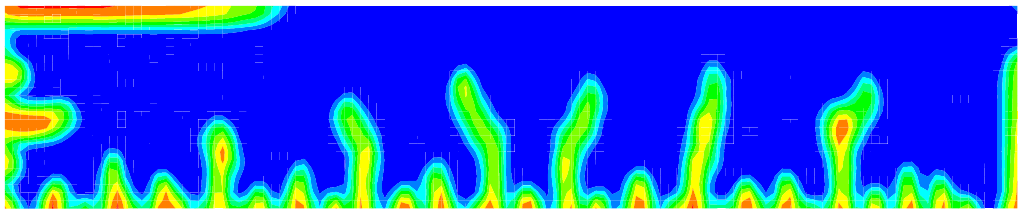}
}
	\subcaptionbox{$T_0=600^\circ C$.}{
	\includegraphics[width=0.3\linewidth]{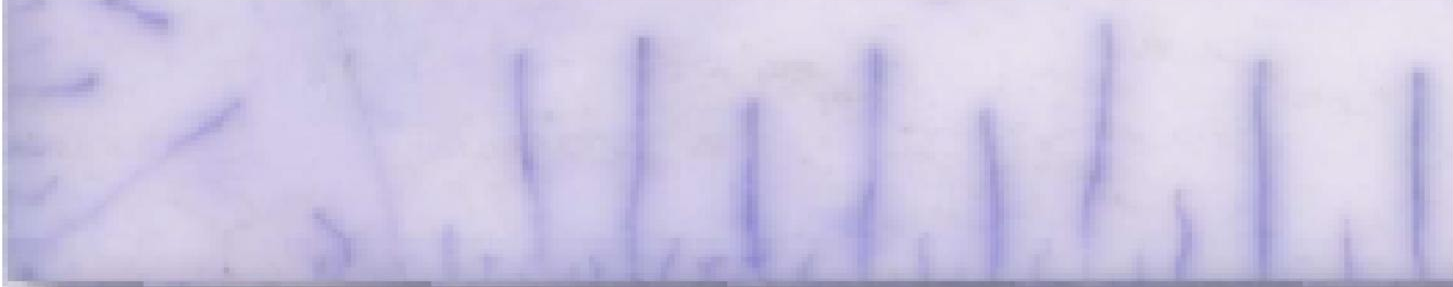}
}
	\caption{Final crack patterns obtained by the proposed model and the experiment\cite{li_non-local_2015}.}
	\label{fig:ceramics}
\end{figure}

\begin{figure}[h!]
	\centering
	\subcaptionbox{$T_0=300^\circ C$.}{
		\includegraphics[width=0.45\linewidth]{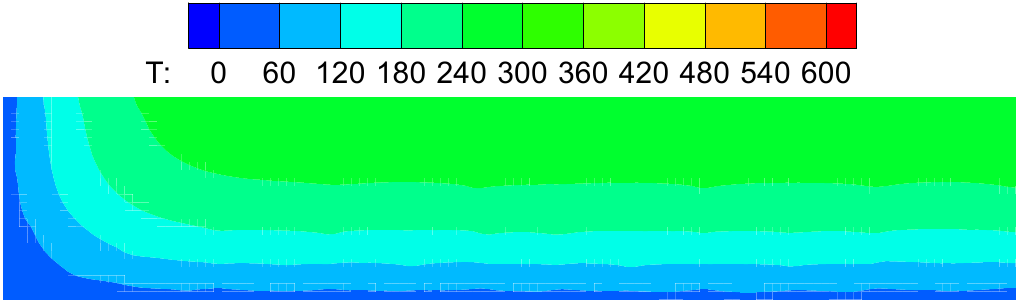}
	}
	\subcaptionbox{$T_0=350^\circ C$.}{
		\includegraphics[width=0.45\linewidth]{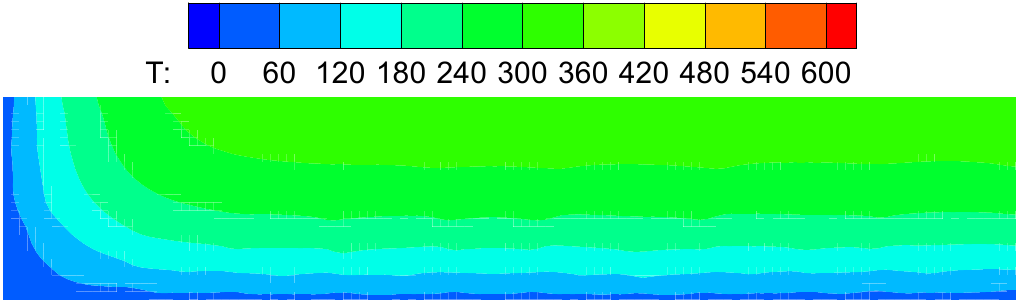}
	}
	\subcaptionbox{$T_0=400^\circ C$.}{
		\includegraphics[width=0.45\linewidth]{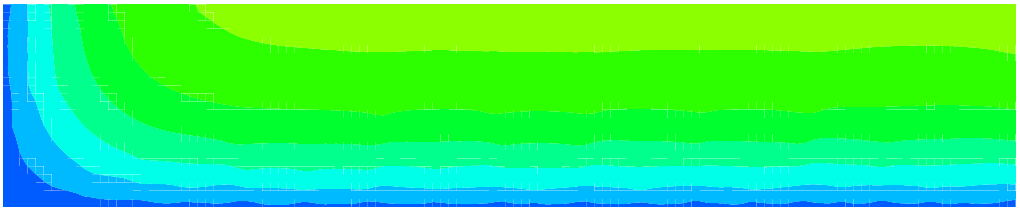}
	}
	\subcaptionbox{$T_0=500^\circ C$.}{
		\includegraphics[width=0.45\linewidth]{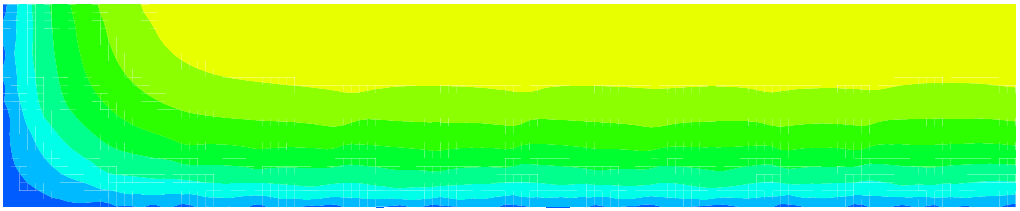}
	}
	\subcaptionbox{$T_0=600^\circ C$.}{
		\includegraphics[width=0.45\linewidth]{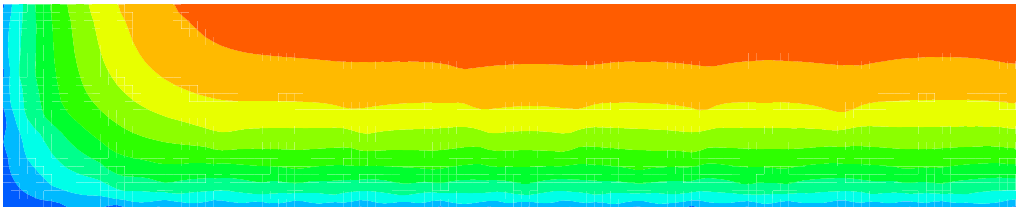}
	}
	\caption{Final temperature field contour obtained by the proposed model.}
	\label{fig:ceramicsT}
\end{figure}

\begin{figure}[h!]
	\centering
	\includegraphics[width=0.6\linewidth]{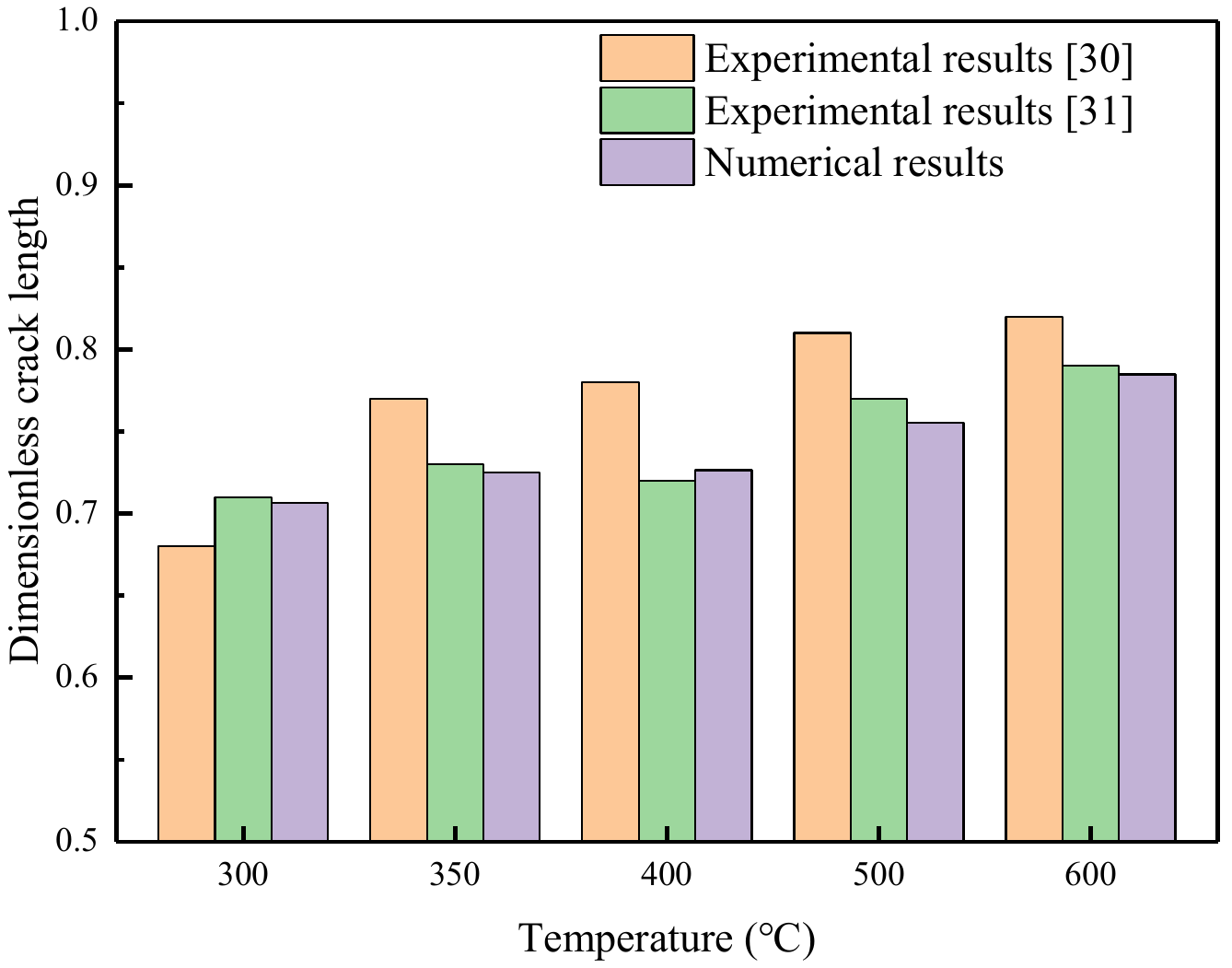}
	\caption{Comparisons between the numerical and experimental results\cite{jiang_study_2012,li_non-local_2015}.}
	\label{fig:cracklength}
	
\end{figure}

Ceramic materials exhibit excellent mechanical properties at high temperatures, but they may break when subjected to sudden temperature changes. Consequently, thermal shock resistance is a crucial metric for assessing the suitability of ceramic materials for high-temperature engineering applications. The quenching test of ceramics is a widely adopted method to investigate the failure mechanisms associated with thermal shock-induced crack patterns. In the previous experiments, thin specimens measuring $50 \times 10 \times 1$mm were heated to temperature $T_0$ and then rapidly quenched in a water bath maintained at $T=20 ^\circ C$ \cite{jiang_study_2012}. Due to the central symmetry of the boundary conditions and geometric model, only a quarter of the numerical model, with dimensions of $25 \times 10$mm, needs to be simulated. Fig. \ref{fig:fig9} presents a simplified two-dimensional numerical simulation sketch, illustrating the geometry and boundary conditions of the computational domain. The upper and right edges are constrained, while convection boundaries are applied to the left and bottom edges. The numerical computations are based on the plane stress assumption. Table \ref{tab:tb4} lists the material properties reported in \cite{li_non-local_2015,sun2021}. A uniform finite element discretization with a size of $0.1 \times 0.1$mm is employed to calculate both the temperature and displacement fields. The PD horizon is set to $\delta = 3\Delta x$. Fig. \ref{fig:ceramics} compares the final crack patterns obtained through numerical simulations and experiments \cite{li_non-local_2015}. Both methods demonstrate crack propagation under thermal shock conditions. The initial temperature $T_0$ and convective heat transform coefficienct $h_s$ of the ceramic plates influence the final crack patterns. The left and middle columns of the figure show the new definition of the PD damage $\hat{\boldsymbol d}$ proposed in this paper for different $T_0$ values. The right column displays the final crack patterns obtained from experiments \cite{li_non-local_2015}. Fig. \ref{fig:ceramicsT} presents the final temperature field contours obtained through numerical simulations for various initial temperatures. To compare numerical simulations with experiments, the dimensionless crack length is selected as a reference index. The dimensionless crack length is defined as the ratio of the crack length to the total height (5mm). Cracks with a dimensionless crack length greater than 0.6 are classified as long cracks. Fig. \ref{fig:cracklength} compares the numerical results from the proposed model with experimental results. The crack length in numerical simulations is slightly shorter than that observed in experiments. This discrepancy may be attributed to unpredictable manufacturing defects or material heterogeneities in the ceramic specimens used in the experiments.

\begin{table}[h!]
	\caption{\textbf{Material properties of ceramics}}
	\centering
	\begin{tabular}{ccc}
		\toprule
		Parameter&Value&Unit\\
		\midrule
		Young's modulus $E$&370e9&$Pa$\\
		Poisson’s ratio $\nu$ &0.33&–\\
		Density $\rho$&3980&$kg/m^3$\\
		Thermal conductivity $k$&31&$J/(s\cdot m\cdot K)$\\
		Specific heat capacity $c$&880&$J/(kg \cdot K)$\\
		Thermal expansion coefficient $\alpha$&7.5e-6&$1/K$\\
		Fracture energy $G$&42.47&$N/m$\\
		\multirow{5}{*}{Convective heat transfer coefficient $h_s$}&65,000 ($T_0 = 300^\circ C$)&\multirow{5}{*}{$W/m^2 \cdot K$}\\&90,000 ($T_0 = 350^\circ C$)&\\&82,000 ($T_0 = 400^\circ C$)&\\&70,000 ($T_0 = 500^\circ C$)&\\&60,000 ($T_0 = 600^\circ C$)&\\
		\bottomrule
	\end{tabular}
	\label{tab:tb4}
\end{table}

\clearpage
\section{Conclusion}\label{sec:sec6}
A new definition of PD damage has been developed to model thermal fractures. A comprehensive solution framework has been proposed to compute both the temperature and displacement fields. Specifically, the temperature field is calculated by CCM, whereas the displacement field is computed via PD. To ensure accuracy, a unified finite element discretization is employed for both methods. To elucidate the impact of thermal cracks or defects on the temperature field, the insulation crack is modeled by reducing thermal conductivity through the new definition of the PD damage, which indicates the influence of various bonds distributions on thermal conductivity.

The proposed thermo-mechanical model utilizes finite element discretization. The shared mesh system for both temperature and displacement computation eliminates the need for remeshing, making it highly efficient and convenient for engineering applications. Furthermore, the new definition of the PD damage differs from original definition. The original definition can only represent the number of bond failure within the domain and can't depict specific bond failure distribution. In contrast, the new definition of the PD damage can characterize damage in different directions. Notably, the model presented in this paper is applicable not only to isotropic materials but also to anisotropic materials. For isotropic materials, the new PD damage is defined along the three coordinate axis directions, whereas for anisotropic materials, it is defined along the material's three principal directions. Subsequent researches will provide a discussion.
\section*{Acknowledgements}\label{sec:sec7}
The authors gratefully acknowledge the financial support received from the National Natural Science Foundation of China, PRChina (12272082).
\clearpage
\bibliographystyle{elsarticle-num} 
\bibliography{refer}

\end{document}